%% file: manuscript.tex
\documentclass[12pt]{article}

\usepackage[T1]{fontenc}
\usepackage[utf8]{inputenc}
\IfFileExists{lmodern.sty}{\usepackage{lmodern}}{\usepackage{mathptmx}}
\usepackage{amsmath,amssymb,amsthm}
\usepackage{graphicx}
\usepackage{xcolor}
\usepackage{booktabs,longtable,array,multirow}
\usepackage{pdflscape}
\usepackage{caption}
\usepackage{float}
\usepackage{enumitem}
\usepackage[margin=1in]{geometry}
\usepackage{microtype}
\usepackage{natbib}
\usepackage[hyphens]{url}
\usepackage[colorlinks=true,linkcolor=blue,citecolor=blue,urlcolor=blue]{hyperref}
\usepackage{setspace}
\usepackage[most]{tcolorbox}
\newtcolorbox{mathbox}[1]{breakable,enhanced,colback=black!4,colframe=black!35,boxrule=0.4pt,arc=2pt,
  left=6pt,right=6pt,top=4pt,bottom=4pt,fontupper=\small,title={#1},fonttitle=\bfseries\small,
  coltitle=black,colbacktitle=black!12,before skip=10pt,after skip=10pt}
\newtcolorbox{keybox}[1]{breakable,enhanced,colback=blue!4,colframe=blue!45!black,boxrule=0.6pt,arc=2pt,
  left=8pt,right=8pt,top=6pt,bottom=6pt,title={#1},fonttitle=\bfseries,coltitle=white,colbacktitle=blue!45!black}

\usepackage{xr}
\IfFileExists{agsm.bst}{\bibliographystyle{agsm}}{\bibliographystyle{plainnat}}
\setcitestyle{authoryear,round,aysep={},yysep={,}}

\hypersetup{
  pdftitle={Claims, Designs, and Guarantees: A Critical Narrative Review of Statistical Inference Reform and Practice, 2016--2026},
  pdfauthor={Aswini Kumar Patra; Shashi Bhushan Jha},
  pdfkeywords={p-value, compatibility interval, S-value, equivalence testing, second-generation p-value, Bayes factor, e-value, anytime-valid inference, false discovery rate, open science, reporting guidelines}}

\newcolumntype{L}[1]{>{\raggedright\arraybackslash}p{#1}}
\setlist{itemsep=2pt,topsep=4pt}
\newtheorem{proposition}{Proposition}
\theoremstyle{definition}

\newcommand{\showauthor}{1}

\begin{document}

\title{\bfseries Claims, Designs, and Guarantees: A Critical Narrative Review of Statistical Inference Reform and Practice, 2016--2026}
\if1\showauthor
  \author{Aswini Kumar Patra$^{1,}$\thanks{Corresponding author. Department of Computer Science \& Engineering, North Eastern Regional Institute of Science \& Technology, Nirjuli, Itanagar, Arunachal Pradesh, India.}\ \ and\ \ Shashi Bhushan Jha$^{2}$\\[8pt]
  \small $^{1}$Department of Computer Science \& Engineering,\\ \small North Eastern Regional Institute of Science \& Technology, Arunachal Pradesh, India\\[2pt]
  \small $^{2}$Department of Electrical Engineering and Computer Science,\\ \small Embry-Riddle Aeronautical University, Daytona Beach, FL, USA}
\else
  \author{}
\fi
\date{}
\maketitle

\begin{abstract}
\noindent
Statistics helps scientists decide what their data show, but its most familiar tool, the $p$-value, is widely misunderstood. Between 2016 and 2026, professional bodies issued new guidance, and researchers developed or popularised alternatives and complements: compatibility intervals, equivalence tests, Bayes factors, and e-values that remain valid when data are checked repeatedly. Reporting rules for trials and reviews were also rewritten. Yet practice has changed little: in the biomedical literature, about 96\% of papers that report $p$-values still include at least one below 0.05. This review argues that much of the confusion comes from treating these numbers as interchangeable. Each inferential procedure carries an intended interpretation and, where applicable, a formal guarantee, such as a 5\% false-alarm rate, that holds only if the study was run and analysed in a particular way. We call this the Claim--Design--Guarantee framework: a result licenses a claim only when the claim, the design (assignment, sampling, monitoring, selection, and analysis), and the guarantee of the method are aligned. We trace the main developments of the decade, explain each tool in plain terms with formal details in boxes, and show with reproducible examples when the choice of tool changes the conclusion. For example, in a simulated two-group study with 100 interim looks, repeatedly applying an unadjusted $p<0.05$ rule produced false positives in 37.6\% of studies, whereas the particular anytime-valid e-value procedure we used produced 2.2\%, below its theoretical 5\% upper bound. We end with a decision tree, a method-selection table, and a step-by-step workflow linked to current reporting guidelines.

\end{abstract}

\newpage
\begin{keybox}{Plain-language summary}
When a study reports that a result is ``statistically significant,'' it usually means a number called the $p$-value fell below 0.05. Many people read this as proof that an effect is real or important. It is neither. A $p$-value only says how surprising the data would be if there were no effect, and an ordinary $p$-value keeps that meaning only if the data were collected and analysed in the way it assumes---for example, without repeatedly checking results and stopping when they look good. Over the past ten years statisticians have offered clearer ways to report results: giving a range of plausible effect sizes, deciding in advance how large an effect must be to matter, weighing the evidence for competing explanations, and using methods that stay trustworthy even when researchers look at their data many times. This review explains these tools without assuming a statistics background, shows with simple examples when they lead to different conclusions, and gives practical steps for choosing among them. Its central message is that every inferential result should come with two pieces of information: what claim it supports, and under what conditions it can be trusted.
\end{keybox}

\begin{keybox}{Key points}
\begin{itemize}[leftmargin=*]
\item A $p$-value is not the probability that a hypothesis is true, and ``not significant'' does not mean ``no effect.''
\item Every inferential procedure carries an intended interpretation and, where applicable, a formal guarantee (such as a 5\% false-alarm rate or 95\% coverage) that holds only under the design for which it was built.
\item Looking at the data repeatedly and stopping at the first unadjusted $p<0.05$ inflates false alarms (to 38\% with 100 looks in our simulation); methods built for repeated looks, such as group-sequential designs and e-values, keep them controlled.
\item The same $p$-value can accompany a large effect in a small study and a trivially small effect in a very large one; report the effect size and its range.
\item To claim an effect is negligible, decide in advance what ``too small to matter'' means and use an equivalence test; evidence for an exactly zero effect is a different claim.
\item Pre-registering the plan and following reporting guidelines (SPIRIT 2025, CONSORT 2025, PRISMA 2020) matter as much as the choice of statistical method.
\end{itemize}
\end{keybox}

\newpage
\onehalfspacing

\section{Introduction}\label{sec:intro}

Most quantitative research ends with a statement about whether an effect is ``real.'' The usual tool is \emph{null-hypothesis significance testing} (NHST): the researcher assumes there is no effect (the \emph{null hypothesis}), computes a \emph{$p$-value}---the probability of seeing data at least as extreme as those observed if that assumption and all others behind the analysis were true---and calls the result \emph{statistically significant} if $p$ falls below a threshold, usually 0.05. NHST persists because it is a compact convention, but its routine use has encouraged researchers to treat ``statistical significance'' as a proxy for truth, importance, or publishability \citep{Goodman1999,Greenland2016,Wasserstein2016}. The decade examined here opened with the American Statistical Association (ASA) statement on $p$-values \citep{Wasserstein2016}, which followed a journal's decision to ban null-hypothesis testing procedures altogether \citep{TrafimowMarks2015} and large \emph{replication} projects---in which independent teams repeat published studies to see whether the findings hold---that reported lower-than-expected success rates \citep{OSC2015}. It continued through proposals to lower the default threshold to 0.005 \citep{Benjamin2018}, counter-proposals to justify the threshold (written $\alpha$) case by case \citep{Lakens2018JustifyAlpha}, calls to abandon or retire statistical significance \citep{McShane2019,Amrhein2019}, a special issue of \emph{The American Statistician} with 43 papers \citep{Wasserstein2019}, and a 2021 ASA President's Task Force statement emphasising that $p$-values and significance tests, properly applied and interpreted, remain valid tools \citep{Benjamini2021TaskForce,Kafadar2021}.

Over the same period three methodological strands matured. The first reframed classical outputs. $P$-values and \emph{confidence intervals} (ranges of effect sizes consistent with the data) were recast as measures of \emph{compatibility} between the data and the full set of assumptions, and \emph{S-values} re-expressed $p$ as ``bits of surprise'' \citep{Greenland2019,Rafi2020,Mansournia2022,Gelman2019CI}. The second made \emph{practical relevance} explicit. Researchers specify a \emph{smallest effect size of interest} (SESOI), the smallest effect worth caring about, and then use \emph{equivalence tests} to show an effect is smaller than that \citep{Lakens2017,Lakens2018Tutorial} or \emph{second-generation $p$-values} to measure how much of the plausible range is negligible \citep{Blume2018,Blume2019}. The third addressed the \emph{process} of analysis. Bayesian methods, which express uncertainty as probabilities updated by data, gained a disciplined ``workflow'' and reporting guidelines \citep{Gelman2020Workflow,Kruschke2021BARG}. \emph{E-values} were developed as measures of evidence that stay valid even when researchers check their data repeatedly and decide when to stop (\emph{optional stopping}) \citep{Vovk2021,Shafer2021,Howard2021,Ramdas2023SAVI,Grunwald2024,RamdasWang2025}. Structural reforms proceeded in parallel: \emph{preregistration} (publicly recording the study plan before collecting data), \emph{Registered Reports} (journals accepting a study on its plan, before results are known), \emph{multiverse analysis} (showing results under all reasonable analysis choices), and updated reporting guidelines \citep{Munafo2017,Nosek2018,Steegen2016,Chambers2022,Page2021,Chan2025SPIRIT,Hopewell2025}.

Practice, however, has changed less than guidance. \emph{Meta-research}, the study of how research is done, makes this visible. A text-mining study of PubMed abstracts and PubMed Central full texts from 1990 to 2015 found that $p$-value reporting was rising and that 96\% of abstracts reporting $p$-values contained at least one $p\le 0.05$ \citep{Chavalarias2016}. A preprint extending the same analysis to 2025 reported that the proportion of papers with at least one $p\le 0.05$ has remained between 94\% and 98\% since 1998, with continued clustering at 0.05 and 0.001 \citep{Choi2026}. Comparing standard psychology articles with Registered Reports, \citet{Scheel2021} found positive results for the first hypothesis in 96\% versus 44\% of articles. An analysis of 23{,}551 randomised trials in the Cochrane Database estimated that, for the $p$-values most often reported as significant, effect estimates are exaggerated on average and intervals labelled ``95\%'' cover the true effect only about 90\% of the time \citep{vanZwet2024}.

\paragraph{What is missing.} Existing reviews have covered parts of this landscape well: the relationship between $p$-values and Bayes factors \citep{HeldOtt2018}, misinterpretations of classical outputs \citep{Greenland2016}, uncertainty reporting in economics \citep{Imbens2021}, broad statistical ideas of the past half-century \citep{Gelman2021Ideas}, and the theory of anytime-valid inference \citep{Ramdas2023SAVI,RamdasWang2025}. What practitioners still lack is a single account that (a) places classical, relevance-based, Bayesian, and e-value methods on a common footing; (b) states for each \emph{which claim} it supports and \emph{under which design conditions} its guarantee holds; (c) demonstrates, rather than asserts, when method choice changes conclusions; and (d) connects these choices to the reporting standards that journals and regulators now enforce.

\begin{keybox}{A running example: does a reading app help children read better?}
Throughout this review we follow one simple study. A school district wants to know whether a reading app, used for 15 minutes a day, improves children's reading. Pupils are randomly divided into two groups: one uses the app for a term, the other continues as usual. At the end of term all pupils sit the same reading test, and the difference between the groups' average scores is expressed in \emph{standard-deviation units}, a common scale on which 0.2 is conventionally regarded as a small effect. Before the study, the district decides that a gain smaller than 0.2 would not justify the cost of licences and teacher time. The district then has to answer several different questions: How large is the gain? Is there any gain at all? Is the gain too small to matter? Should we buy the app? Each question needs a different statistical tool, and the answer can change depending on how the study was run. We return to this example in every section.
\end{keybox}

\paragraph{How to read this review.} Technical terms are defined when first used and collected in a glossary (Supplementary Appendix~\ref{app:glossary}). Formal definitions appear in shaded boxes that can be skipped without losing the argument.

\paragraph{Contributions.} This review makes four contributions. (1) It proposes the Claim--Design--Guarantee (CDG) framework (Section~\ref{sec:cdg}), which extends the familiar distinction between evidence-centric and decision-centric inference by making design conditions explicit. (2) It synthesises the 2016--2026 literature in a chronology and a structured comparison of tools that records each tool's guarantee, failure conditions, and characteristic misreading (Sections~\ref{sec:decade}--\ref{sec:tools}). (3) It provides reproducible demonstrations---a worked two-arm example and simulations---of the regimes in which method choice does and does not alter conclusions (Section~\ref{sec:matter}). (4) It translates the synthesis into a method-selection matrix, a decision tree, and a three-phase workflow mapped to SPIRIT 2025, CONSORT 2025, PRISMA 2020, the Bayesian Analysis Reporting Guidelines (BARG), and SAMPL (Sections~\ref{sec:matrix}--\ref{sec:workflow}).

\section{Scope and approach}\label{sec:scope}

\paragraph{Review type and rationale.} This is a critical narrative review. A systematic review would be the right design for a narrow, well-defined question, such as the prevalence of a reporting practice in a defined literature. The question here is broader and conceptual: how the tools and norms of statistical inference changed over a decade, and how they relate to one another. That question requires synthesis across statistics, methodology, meta-research, and regulatory guidance, and judgement about which contributions mattered. We therefore followed a structured narrative approach and report it against the six criteria of the Scale for the Assessment of Narrative Review Articles (SANRA): justification of importance, statement of aims, description of the literature search, referencing, scientific reasoning, and presentation of data \citep{Baethge2019SANRA}.

\paragraph{Aims.} The review has three aims: (i) to synthesise the main developments in statistical inference and its practice between 2016 and 2026; (ii) to clarify, for each major tool, which claims it can support and under which study conditions its guarantees hold; and (iii) to translate this synthesis into practical guidance for researchers who apply statistics.

\paragraph{Time frame.} The core window runs from 2016, when the ASA statement on $p$-values was published, to October 2026. Earlier works are included only where they define a concept still in active use, such as the Neyman--Pearson framework, the Benjamini--Hochberg procedure, group-sequential boundaries, equivalence testing, and the Jeffreys--Lindley paradox.

\paragraph{Sources and search.} Literature was identified from four complementary sources: (a) bibliographic databases (Web of Science, Scopus, PubMed, and MathSciNet), searched with tool-specific terms, for example ``equivalence test'' AND ``smallest effect size of interest,'' ``second-generation p-value,'' ``e-value'' AND (``anytime-valid'' OR ``safe testing''), and ``Bayes factor'' AND ``optional stopping''; (b) the official statements of the American Statistical Association and the 43 papers of the 2019 special issue of \emph{The American Statistician}, with backward and forward citation tracking from each \citep{Wasserstein2016,Wasserstein2019,Benjamini2021TaskForce}; (c) the EQUATOR Network library and the websites of the International Council for Harmonisation, for reporting and regulatory guidance; and (d) the tables of contents of leading statistics and methodology journals over the review period.

\paragraph{Eligibility.} Works were included if they (i) introduced or substantially clarified an inferential method or its guarantees; (ii) constituted official guidance from a professional, regulatory, or reporting-guideline body; or (iii) provided large-scale empirical evidence on statistical practice, replicability, or analytic variability. Applications of established methods to a single substantive question, software documentation, and opinion pieces without methodological or empirical content were excluded.

\paragraph{Source hierarchy.} Peer-reviewed journal articles, books, and official guidance were the primary evidence base and support every substantive claim in the review. Apart from books and official guidance, only three cited works have not been through journal peer review: the Bayesian workflow paper \citep{Gelman2020Workflow}, which is widely used and has no journal version, and is supplemented here by a peer-reviewed textbook and articles on the same practices \citep{Gelman2013BDA,Vehtari2017}; a 2026 meta-research update on $p$-value reporting \citep{Choi2026}, which extends a peer-reviewed study by the same group using the same methods \citep{Chavalarias2016} and is used only to indicate that the earlier pattern persists; and the ICH E20 guideline on adaptive designs, which was released as a draft for public consultation \citep{ICHE20}. Each is labelled as such where it is cited, and no conclusion rests on it alone. 

\paragraph{Synthesis and verification.} Included works were grouped by tool and by type of contribution (methods, guidance, or empirical evidence), and synthesised narratively around the framework introduced in Section~\ref{sec:cdg}. Bibliographic details were verified against publisher records and Crossref. Empirical findings are reported with their study design and setting so that readers can judge their weight; no formal risk-of-bias assessment was undertaken, which is a limitation discussed in Section~\ref{sec:discussion}. All numerical illustrations are reproducible from the provided supplementary code.
\section{The Claim--Design--Guarantee framework}\label{sec:cdg}

\subsection{Two goals of inference}\label{sec:goals}

The distinction between \emph{evidence-centric} and \emph{decision-centric} inference remains useful. \emph{Inference} means drawing conclusions about a wider population or process from limited data. Evidence-centric inference asks what the data, under stated assumptions, indicate about the quantities of interest (\emph{parameters}, such as a treatment effect) or about competing explanations; its natural outputs are effect estimates with uncertainty intervals, likelihood ratios, posterior distributions, Bayes factors, and predictive checks \citep{Royall1997,Gelman2013BDA,Mansournia2022}. Decision-centric inference asks which action to take given uncertainty; it requires an explicit specification of actions, losses or utilities, and the error rates one is willing to tolerate \citep{NeymanPearson1933,Lakens2018JustifyAlpha,Grunwald2024PNAS}. The two traditions correspond loosely to Fisher's evidential use of the $p$-value and Neyman and Pearson's error-rate framework, which are often conflated in practice \citep{Lehmann1993,HubbardBayarri2003}.

This dichotomy, however, does not say when a given number may be trusted. A 95\% confidence interval is evidence-centric, but it is no longer a 95\% interval if it is reported only because it excluded zero among many candidates. A $p<0.05$ decision rule is decision-centric, but its 5\% error rate disappears if the analyst keeps collecting data until $p$ crosses the threshold. What matters is the \emph{guarantee} the summary carries and whether that guarantee survives the \emph{design}. The CDG framework makes this explicit.

\subsection{Claims, designs, and guarantees}

\begin{figure}[t]
\centering
\includegraphics[width=\textwidth]{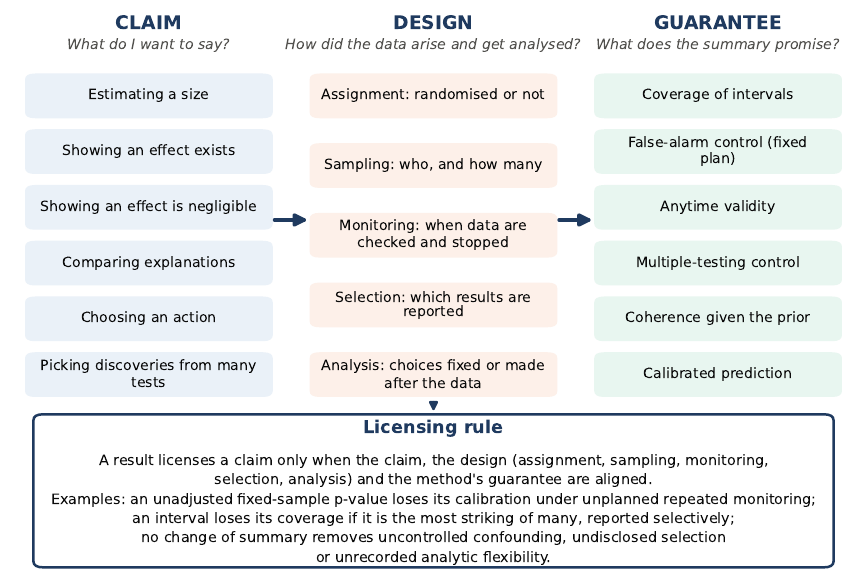}
\caption{The Claim--Design--Guarantee (CDG) framework. A result licenses a claim only when the claim, the design (assignment, sampling, monitoring, selection, and analysis), and the guarantee of the method are aligned.}
\label{fig:cdg}
\end{figure}

Figure~\ref{fig:cdg} summarises the framework. In everyday terms it asks three questions of every reported number: \emph{What am I claiming?} (the claim), \emph{How were the data collected, checked, and analysed?} (the design), and \emph{What does this number promise, and under which conditions?} (the guarantee).

\paragraph{Claims.} We distinguish six kinds of claim, illustrated with the reading-app study.
\begin{itemize}
\item \textbf{Estimating a size:} ``the app raises scores by about 0.3 standard deviations.''
\item \textbf{Showing an effect exists:} ``the app raises scores'' (the direction is positive, whatever the size).
\item \textbf{Showing an effect is negligible:} ``any gain is too small to matter.''
\item \textbf{Comparing explanations:} ``the data are better predicted by a model in which the app has some nonzero effect than by a model in which it has exactly none.''
\item \textbf{Choosing an action:} ``the district should buy the app.''
\item \textbf{Picking discoveries from many tests:} ``of 20 outcomes measured, the app improved spelling and vocabulary.''
\end{itemize}
The first four are evidence claims; choosing an action is a decision; picking discoveries combines both.

\paragraph{Designs.} ``Design'' here means the whole process that produces a reported result, not only the experimental layout. We decompose it into five components, because each affects the validity or scope of a result in a different way:
assignment, sampling, monitoring, selection, and analysis.
\begin{itemize}
\item \textbf{Assignment:} how units come to receive the exposure---by randomisation or by choice or circumstance. Comparing schools that volunteered for the app with schools that did not invites \emph{confounding}: the groups may differ in other ways, such as teacher enthusiasm.
\item \textbf{Sampling:} who is studied and how many. Unlike the other four components, sampling rarely breaks a guarantee outright: it determines the population to which a result applies (external validity) and how precise the estimates are, and it matters most when the sample size is itself chosen in response to the data, which is a monitoring problem.
\item \textbf{Monitoring:} when and how often the data are examined, and the rule for stopping or continuing---for example, checking scores every week and stopping when they look good.
\item \textbf{Selection:} which outcomes, subgroups, and analyses are reported, for example highlighting the most striking of many outcomes.
\item \textbf{Analysis:} whether choices such as exclusions, covariates, transformations, and models were fixed in advance or made after seeing the data \citep{Simmons2011,Steegen2016}.
\end{itemize}
Assignment, monitoring, selection, and analysis can each invalidate a specific guarantee: non-random assignment undermines causal interpretation, unplanned monitoring breaks fixed-sample false-alarm control, selective reporting breaks coverage and error rates, and undisclosed analytic choices do both. Sampling chiefly limits the scope and precision of a valid result. A study with a \emph{fixed plan}---assignment by randomisation, a predetermined sample, no unplanned looks, complete reporting, and a prespecified analysis---is the setting most classical methods assume.

\paragraph{Guarantees.} Every inferential procedure carries an intended interpretation and, where applicable, a formal statistical guarantee. (Purely descriptive statistics, such as a sample mean reported without inference, need not carry one.) We distinguish six kinds of guarantee.
\begin{itemize}
\item \textbf{Coverage:} a method for 95\% intervals captures the true value in 95\% of repeated studies run to a fixed plan.
\item \textbf{False-alarm control under a fixed plan:} a test declares an effect when there is none (a Type~I error) no more often than the chosen rate, provided the plan was fixed in advance.
\item \textbf{Anytime validity:} the same false-alarm or coverage guarantee holds however often the data are checked and whenever the study stops.
\item \textbf{Multiple-testing control:} the false discovery rate (the expected share of false positives among reported findings) or the family-wise error rate (the chance of any false positive) stays below a set level.
\item \textbf{Coherence given the prior:} Bayesian probabilities are internally consistent given the model and prior, but are only as good as those inputs.
\item \textbf{Calibrated prediction:} predictions perform as stated on new data.
\end{itemize}

\begin{keybox}{The licensing rule}
A result licenses a claim only when three things are aligned: the \emph{inferential target} (the claim), the \emph{data-generating and analysis process} (assignment, sampling, monitoring, selection, and analysis), and the \emph{guarantee} of the method used.
\end{keybox}

Three consequences follow directly, again using the reading-app study. First, suppose the district checks the scores every week and stops as soon as $p<0.05$. An ordinary fixed-sample $p$-value retains its advertised calibration only when the sampling and analysis procedure for which it was constructed is respected; unadjusted repeated testing with data-dependent stopping changes that calibration, so the result can no longer support the claim that the app works. Procedures built for monitoring---group-sequential tests, sequentially adjusted or always-valid $p$-values, and e-values (Sections~\ref{sec:evalues} and \ref{sec:gsd})---retain their guarantees under the monitoring they were designed for. Second, if the district measures 20 outcomes and reports the one with the most impressive interval, that interval no longer has its stated coverage. Third, if the app was used only in schools that volunteered, no statistical promise removes the possibility that those schools differed in other ways; claims about cause and effect then need explicit assumptions and sensitivity analysis \citep{HernanRobins2020,VanderWeeleDing2017,CinelliHazlett2020}. Bayesian probabilities remain coherent given the prior, but whether they also keep false alarms rare under peeking or selective reporting depends on whether the prior and model are themselves sound \citep{deHeide2021,Hendriksen2021}.

\subsection{How the framework differs from existing guidance}

The ASA statements emphasise interpretation \citep{Wasserstein2016,Benjamini2021TaskForce}; the CDG framework adds an explicit check of whether the conditions for that interpretation hold. Reporting guidelines specify what to report \citep{Lang2015,Hopewell2025}; the framework explains why each item matters for the validity of a particular claim. The estimand framework of ICH E9(R1) \citep{ICHE9R1} already requires the target of estimation to be defined before analysis; CDG generalises this discipline to the \emph{claim} and the \emph{guarantee}. Finally, the framework is deliberately pluralistic: it does not rank frequentist, Bayesian, or e-value methods but asks whether the chosen method's guarantee matches the claim and the design. What distinguishes it from a general checklist of assumptions is that it locates each failure precisely: a result can fail because the claim is of the wrong type for the method (for example, using a point-null test to claim negligibility), because a component of the design breaks the guarantee (for example, unplanned monitoring), or because the guarantee itself does not cover the claim (for example, coverage does not imply causation).

\section{The decade in review, 2016--2026}\label{sec:decade}

\begin{figure}[p]
\centering
\includegraphics[width=\textwidth]{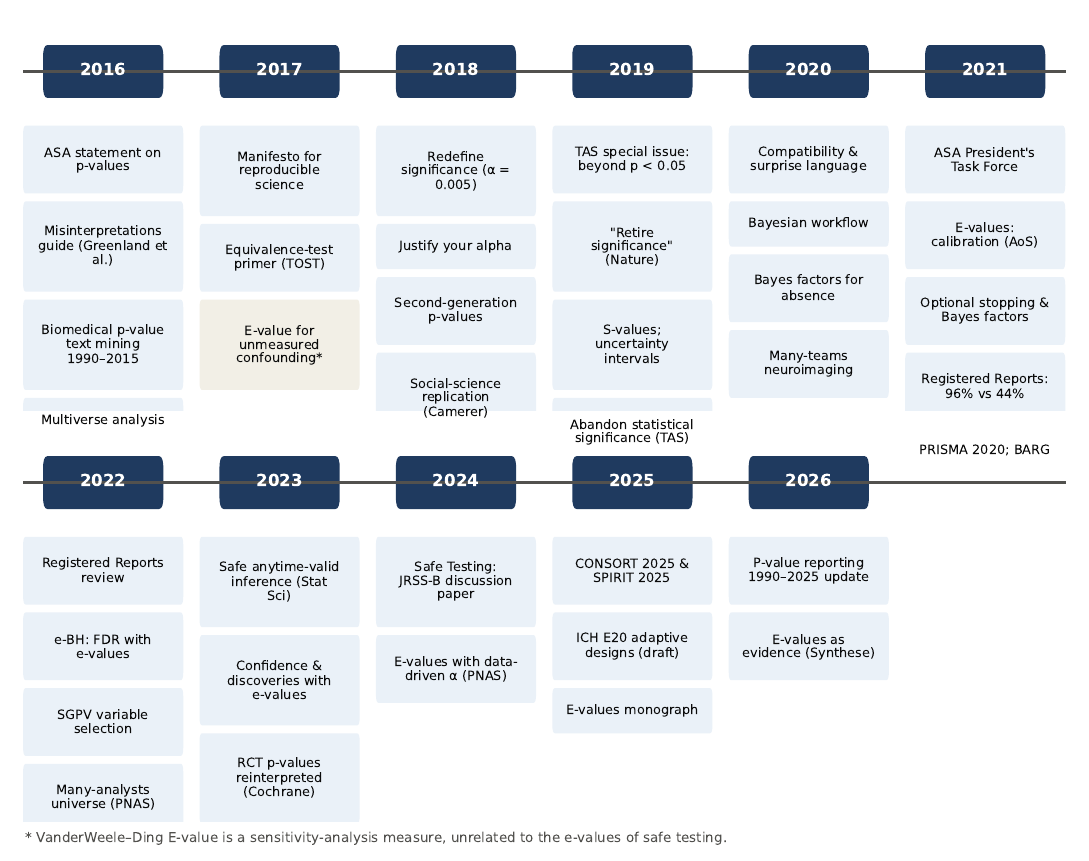}
\caption{Key developments shaping statistical inference practice, 2016--2026. Details and references are given in Table~\ref{tab:chronology}.}
\label{fig:timeline}
\end{figure}

Figure~\ref{fig:timeline} and Table~\ref{tab:chronology} summarise the main developments. We group them into five strands.

\subsection{Statements, debates, and thresholds}

The 2016 ASA statement set out six principles, including that $p$-values indicate how incompatible data are with a specified model, that they do not measure the probability that a hypothesis is true, and that scientific conclusions should not rest only on whether a $p$-value crosses a threshold \citep{Wasserstein2016}. A companion guide catalogued common misinterpretations of tests, intervals, and power \citep{Greenland2016}. Debate then centred on thresholds. \citet{Benjamin2018} proposed redefining significance for claims of new discoveries at $p<0.005$, partly on the grounds that $p\approx 0.05$ corresponds to weak evidence under Bayes-factor calibrations; \citet{Lakens2018JustifyAlpha} argued instead that $\alpha$ should be justified by the costs of errors, prior plausibility, and sample size. Others argued that dichotomisation itself was the problem \citep{McShaneGal2017,McShane2019,Amrhein2019}. The 2019 special issue concluded with the editorial advice not to use the phrase ``statistically significant'' \citep{Wasserstein2019}. In 2021 an ASA President's Task Force, convened in part because the 2019 editorial had been widely read as official ASA policy, affirmed that $p$-values and significance tests, when properly applied and interpreted, strengthen the rigour of conclusions drawn from data, and that thresholds may be appropriate when chosen in context \citep{Benjamini2021TaskForce,Kafadar2021}. The resulting consensus is narrower than either camp's rhetoric: thresholds may be used for decisions when justified, but they should not be equated with scientific importance, and continuous reporting of evidence is preferable for evidential claims \citep{Lakens2021PPS,Imbens2021}.

\subsection{Meta-research on practice and replicability}

Large-scale evidence has shaped the debate. Replication projects in psychology, social science, and preclinical cancer biology reported replication rates and effect sizes well below those of the original studies \citep{OSC2015,Camerer2018,Errington2021}, consistent with earlier analyses of low power and flexible analysis \citep{Ioannidis2005,Simmons2011,Button2013}. Text-mining studies documented that $p$-values became more common and that nearly all papers using them report significant results \citep{Chavalarias2016,Choi2026}. ``Many-analysts'' studies, in which independent teams analysed the same data to answer the same question, found substantial variation in estimates and conclusions that was not explained by team expertise \citep{Silberzahn2018,BotvinikNezer2020,Breznau2022}. These studies are central to the CDG framework because they measure the uncertainty created by analysis choices made after seeing the data, which no single analysis can reveal.

\subsection{Expansion of the inferential toolkit}

Three groups of tools gained traction. \emph{Compatibility-based interpretation} recast $p$-values and intervals as measures of the fit between data and a full model (including all assumptions), with S-values $s=-\log_2 p$ expressing the same information in bits \citep{Greenland2019,Rafi2020,Mansournia2022}. \emph{Relevance-based testing} formalised the SESOI through equivalence and minimum-effect tests \citep{Lakens2017,Lakens2018Tutorial} and through SGPVs, which measure the overlap between an interval estimate and an interval null hypothesis \citep{Blume2018,Blume2019,Zuo2022}. \emph{Bayesian practice} shifted from isolated Bayes-factor tests towards workflows that check whether the prior and model produce sensible data, verify that computations have worked, compare models, and test how sensitive conclusions are to choices \citep{Gelman2020Workflow,Vehtari2017}, with dedicated reporting guidelines \citep{Kruschke2021BARG} and tutorials on using Bayes factors or regions of practical equivalence (ROPE) to support claims of absence \citep{Keysers2020,Kruschke2018ROPE}.

\subsection{Sequential and anytime-valid inference}

The most technically novel development is the theory of e-values and \emph{safe anytime-valid inference} (SAVI): methods whose error guarantees hold no matter how often the data are checked or when the study stops. An e-value can be understood as the winnings of a bettor wagering against the null hypothesis in a game that is fair when the null is true, so large winnings are evidence against it \citep{Shafer2021,Vovk2021,Grunwald2024}; Section~\ref{sec:evalues} explains the idea. Versions updated after every observation (\emph{e-processes}) allow continuous monitoring, and their interval counterparts (\emph{confidence sequences}) are valid at every sample size simultaneously \citep{Howard2021,Ramdas2023SAVI}. The period saw procedures for testing many hypotheses with e-values \citep{WangRamdas2022,Vovk2023}, general constructions for complex hypotheses \citep{Wasserman2020}, industrial use in online experimentation \citep{Johari2022}, living meta-analysis \citep{terSchure2022}, decision-theoretic results showing that e-values support post hoc choices of $\alpha$ \citep{Grunwald2024PNAS}, a monograph \citep{RamdasWang2025}, and a philosophical comparison with likelihoods, Bayes factors, and $p$-values \citep{Chugg2026}. In parallel, work on Bayesian optional stopping showed that the folklore claim that Bayes factors are immune to stopping rules holds only under specific conditions on the prior and hypotheses \citep{deHeide2021,Hendriksen2021}.

\subsection{Open-science infrastructure and reporting standards}

Preregistration and Registered Reports separate confirmatory from exploratory analysis by fixing hypotheses and analysis plans before results are known \citep{Nosek2018,Chambers2022}; the substantially lower rate of positive results in Registered Reports than in standard reports \citep{Scheel2021} is consistent with reduced publication or selective-reporting bias and/or reduced Type~I error inflation, although the observational comparison does not identify which mechanism accounts for the difference. Multiverse and specification-curve analyses make analytic flexibility visible \citep{Steegen2016,Simonsohn2020}. Reporting guidelines were updated: PRISMA 2020 for systematic reviews \citep{Page2021}, SPIRIT 2025 for trial protocols \citep{Chan2025SPIRIT}, and CONSORT 2025 for trial reports, which added an open-science section \citep{Hopewell2025}. For clinical trials, the ICH E9(R1) addendum on estimands \citep{ICHE9R1} and the ICH E20 draft guideline on adaptive designs, released for consultation in June 2025 \citep{ICHE20}, formalise the requirement that adaptive designs (trials whose course can change based on interim results) control the false-alarm rate and that Bayesian designs document their priors and their \emph{operating characteristics}, that is, how often they reach wrong conclusions in simulated scenarios.

\begin{landscape}
\small
\setlength{\tabcolsep}{4pt}
\begin{longtable}{L{1.1cm} L{4.3cm} L{2.2cm} L{6.6cm} L{6.6cm}}
\caption{Chronology of landmark contributions to statistical inference and its practice, 2015--2026 (with selected earlier foundations).}\label{tab:chronology}\\
\toprule
\textbf{Year} & \textbf{Work} & \textbf{Type} & \textbf{Core contribution} & \textbf{Implication for practice}\\
\midrule
\endfirsthead
\toprule
\textbf{Year} & \textbf{Work} & \textbf{Type} & \textbf{Core contribution} & \textbf{Implication for practice}\\
\midrule
\endhead
\midrule \multicolumn{5}{r}{\textit{Continued on next page}}\\
\endfoot
\bottomrule
\endlastfoot
\multicolumn{5}{l}{\textit{Foundations still in use}}\\
1933 & \citet{NeymanPearson1933} & Theory & Most powerful tests; Type~I/II error framework & Basis of decision-centric testing and power analysis\\
1957 & \citet{Lindley1957} & Theory & Fixed $p$ can coexist with Bayes factors favouring $H_0$ as $n$ grows & $p$ and BF diverge in large samples; report both scales with care\\
1979/83 & \citet{OBrienFleming1979}; \citet{LanDeMets1983} & Methods & Group-sequential boundaries; $\alpha$-spending & Planned interim analyses with Type~I error control\\
1987 & \citet{Schuirmann1987} & Methods & Two one-sided tests (TOST) for equivalence & Basis of modern equivalence testing\\
1995 & \citet{BenjaminiHochberg1995} & Methods & False discovery rate and BH procedure & Standard for large-scale discovery\\
2001 & \citet{Sellke2001} & Theory & Bound $-e\,p\ln p$ on evidence against a point null & Translates $p$ into maximum Bayes factor\\
\midrule
\multicolumn{5}{l}{\textit{2015--2019: statements, debates, meta-research}}\\
2015 & \citet{OSC2015} & Meta-research & Large-scale replication of 100 psychology studies & Motivated reform of design and reporting\\
2015 & \citet{TrafimowMarks2015} & Editorial & Journal ban on NHST procedures & Showed demand for alternatives; prompted ASA statement\\
2016 & \citet{Wasserstein2016} & Official statement & Six principles on $p$-values & Foundational guidance on interpretation\\
2016 & \citet{Greenland2016} & Guide & 25 common misinterpretations of tests, intervals, power & Checklist for correct reading of outputs\\
2016 & \citet{Chavalarias2016} & Meta-research & Text mining 1990--2015: rising $p$ reporting; 96\% of abstracts with $p$ report $p\le0.05$ & Quantified selective reporting of significant results\\
2016 & \citet{Steegen2016} & Methods & Multiverse analysis & Makes analytic flexibility visible\\
2017 & \citet{Munafo2017} & Consensus & Manifesto for reproducible science & Agenda spanning methods, reporting, incentives\\
2017 & \citet{Lakens2017} & Tutorial & Equivalence tests for $t$ tests, correlations, meta-analyses & Practical route to claims of negligibility\\
2017 & \citet{VanderWeeleDing2017} & Methods & E-value for unmeasured confounding & Sensitivity analysis for non-randomised studies (unrelated to e-values of SAVI)\\
2018 & \citet{Benjamin2018} & Proposal & Default $p<0.005$ for new discoveries & Threshold debate; evidence calibration\\
2018 & \citet{Lakens2018JustifyAlpha} & Proposal & Justify $\alpha$ from error costs and context & Decision-centric threshold setting\\
2018 & \citet{Blume2018} & Methods & Second-generation $p$-values & Relevance-aware summary with interval null\\
2018 & \citet{Lakens2018Tutorial} & Tutorial & Equivalence testing and SESOI specification & Operational guidance on SESOI\\
2018 & \citet{Camerer2018}; \citet{Silberzahn2018} & Meta-research & Replication of social-science experiments; many analysts, one dataset & Effect inflation and analytic variability\\
2019 & \citet{Wasserstein2019} & Editorial & Special issue: ``beyond $p<0.05$'' & Recommends against dichotomous language\\
2019 & \citet{McShane2019}; \citet{Amrhein2019} & Commentary & Abandon / retire statistical significance & High-visibility reform position\\
2019 & \citet{Greenland2019} & Methods & S-values as information measure & Communicates $p$ on additive bit scale\\
2019 & \citet{Gelman2019CI} & Commentary & ``Uncertainty intervals'' & Highlights model dependence of intervals\\
2019 & \citet{ICHE9R1} & Regulatory & Estimand framework & Requires defining the target before analysis\\
\midrule
\multicolumn{5}{l}{\textit{2020--2023: toolkit, Bayesian practice, anytime-valid inference}}\\
2020 & \citet{Rafi2020} & Methods & Compatibility and surprise language & Reduces misreadings of $p$ and intervals\\
2020 & \citet{Gelman2020Workflow} & Methods & Bayesian workflow & Iterative model building and checking\\
2020 & \citet{Keysers2020} & Tutorial & Bayes factors for evidence of absence & Supports claims of absence in neuroscience\\
2020 & \citet{BotvinikNezer2020} & Meta-research & 70 teams analyse one neuroimaging dataset & Analytic variability alters conclusions\\
2021 & \citet{Benjamini2021TaskForce} & Official statement & ASA President's Task Force & $p$-values valid when properly used; thresholds context-dependent\\
2021 & \citet{Vovk2021}; \citet{Shafer2021} & Theory & E-values: calibration and combination; testing by betting & Foundations of e-value inference\\
2021 & \citet{deHeide2021}; \citet{Hendriksen2021} & Theory & Conditions under which Bayes factors handle optional stopping & Qualifies ``Bayesians can ignore stopping rules''\\
2021 & \citet{Scheel2021} & Meta-research & 96\% vs 44\% positive results (standard vs Registered Reports) & Evidence for selective reporting in standard literature\\
2021 & \citet{Kruschke2021BARG}; \citet{Page2021} & Guidelines & BARG; PRISMA 2020 & Reporting standards for Bayesian analyses and reviews\\
2022 & \citet{WangRamdas2022} & Methods & e-BH procedure & FDR control under arbitrary dependence\\
2022 & \citet{Breznau2022} & Meta-research & 73 teams, same data and hypothesis & ``Hidden universe'' of analytic uncertainty\\
2022 & \citet{Chambers2022} & Review & Registered Reports: evidence and practice & Infrastructure for confirmatory research\\
2023 & \citet{Ramdas2023SAVI} & Review & Game-theoretic statistics and SAVI & Unified account of anytime-valid inference\\
\midrule
\multicolumn{5}{l}{\textit{2024--2026: consolidation and standards}}\\
2024 & \citet{vanZwet2024} & Meta-research & 23{,}551 Cochrane RCTs: exaggeration, coverage, replication by $p$ & Empirical reinterpretation of trial $p$-values\\
2024 & \citet{Grunwald2024}; \citet{Grunwald2024PNAS} & Theory & Safe testing (discussion paper); e-values with data-driven $\alpha$ & Type~I risk control under optional continuation and post hoc losses\\
2025 & \citet{Hopewell2025}; \citet{Chan2025SPIRIT} & Guidelines & CONSORT 2025 and SPIRIT 2025 & Open-science items for trial protocols and reports\\
2025 & \citet{ICHE20} & Regulatory (draft) & Adaptive designs for confirmatory trials & Type~I error control, Bayesian designs, operating characteristics\\
2025 & \citet{RamdasWang2025} & Monograph & Hypothesis testing with e-values & Reference text for e-value methods\\
2026 & \citet{Choi2026} & Meta-research (preprint) & $p$-value reporting 1990--2025: 94--98\% with $p\le0.05$ & Practice has changed little despite reform\\
2026 & \citet{Chugg2026} & Philosophy/theory & E-values compared with likelihoods, BFs, $p$ & Evidential interpretation of e-values\\
\end{longtable}
\end{landscape}
\normalsize

\section{The tools, explained}\label{sec:tools}

This section explains each tool in plain terms: what question it answers, what it guarantees, and how it is most often misread. Software implementing each tool is listed in Supplementary Appendix~\ref{app:software} (Table~\ref{tab:software}). The mathematical definitions are collected in shaded boxes, which readers can skip without losing the argument. Table~\ref{tab:tools} summarises the section.

\subsection{$p$-values: how surprising are the data?}\label{sec:p}

A \emph{$p$-value} answers one narrow question: if there were truly no effect, and every assumption behind the analysis were correct, how often would we see data at least as extreme as ours? A small $p$-value means the data would be surprising under that ``no-effect'' model (the \emph{null hypothesis}). It does not tell us the probability that the null hypothesis is true, the probability that the result is ``due to chance,'' or how large or important the effect is \citep{Goodman1999,Greenland2016}. Because the model includes every assumption---the shape of the data, independence of observations, no selective reporting, and, for an ordinary fixed-sample $p$-value, a predetermined rule for when to stop collecting data---a small $p$ signals that \emph{something} in the model is at odds with the data, not necessarily the no-effect assumption itself \citep{Wasserstein2016}. This is why recent guidance describes $p$ as a measure of \emph{compatibility} between data and model.

When used as a yes/no rule (``reject if $p<0.05$''), the $p$-value carries a precise guarantee: if the null hypothesis is true and the sampling and analysis procedure for which the $p$-value was constructed is respected, the rule raises a false alarm (a \emph{Type~I error}, or false positive) at most 5\% of the time. Sequentially adjusted, group-sequential, and always-valid $p$-values exist for studies that are monitored as data accrue (Sections~\ref{sec:evalues}--\ref{sec:gsd}); the problem addressed in this review is the use of an \emph{ordinary} fixed-sample $p$-value under unplanned repeated monitoring. The threshold, usually written $\alpha$, is a choice, not a law of nature \citep{Lakens2018JustifyAlpha}.

\begin{figure}[t]
\centering
\includegraphics[width=\textwidth]{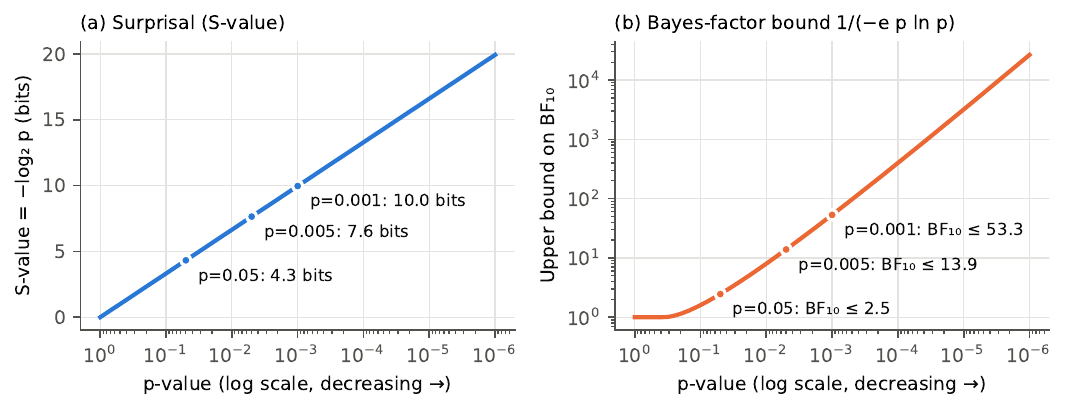}
\caption{Two ways to re-express a $p$-value. (a) The S-value in bits of surprise. (b) The largest Bayes factor in favour of an effect that a given $p$-value can correspond to \citep{Sellke2001}.}
\label{fig:pscales}
\end{figure}

Two re-expressions make $p$ easier to read (Figure~\ref{fig:pscales}). The \emph{S-value} converts $p$ into ``bits of surprise,'' the number of heads in a row from a fair coin that would be equally surprising \citep{Greenland2019,Rafi2020}. A $p$ of 0.05 equals about 4.3 bits---roughly as surprising as four heads in a row, which is not very surprising. A second conversion asks how strongly $p$ could possibly favour an effect over no effect in Bayesian terms: at $p=0.05$ the most favourable Bayes factor (defined in Section~\ref{sec:bayes}) is about 2.5 to 1, and at $p=0.005$ about 14 to 1 \citep{Sellke2001,BenjaminBerger2019}. This bound is why some authors regard $p\approx0.05$ as weak evidence \citep{Benjamin2018}.

\noindent\textit{In the reading-app study.} Suppose the trial of 100 pupils per group (Section~\ref{sec:worked}) gives $p=0.034$. This says that, if the app had no effect and the study had followed its plan, a difference at least this large in either direction would turn up in about 3 out of every 100 such studies. It does \emph{not} say there is a 97\% chance that the app works, and it says nothing about whether the gain is big enough to justify the cost. As an S-value, $p=0.034$ is about 4.9 bits: roughly as surprising as five heads in a row.

\begin{mathbox}{Box 1. $p$-value, S-value, and the Bayes-factor bound}
For a test statistic $T$, where larger values indicate greater departure from the null model $M_0$, the $p$-value for an observed value $t$ is
\[ p(t)=\Pr_{M_0}(T\ge t). \]
For an exact test with a continuous statistic under a fully specified null, the $p$-value is uniformly distributed on $(0,1)$. More generally, a \emph{valid} $p$-value is super-uniform under the null: $\Pr_{H_0}(p\le\alpha)\le\alpha$ for every $\alpha$, with equality not required for discrete or conservative procedures. Rejecting when $p\le\alpha$ therefore gives a Type~I error rate of at most $\alpha$ under the design for which $p$ was constructed \citep{Fisher1925,NeymanPearson1933}.

The S-value is $s=-\log_2 p$ bits \citep{Greenland2019}. For a point null and $p<1/e$, the Bayes factor against $H_0$ is bounded above, over a broad class of alternatives, by $1/(-e\,p\ln p)$ \citep{Sellke2001}; see also \citet{BergerSellke1987}.
\end{mathbox}

\subsection{Confidence intervals: the range of values compatible with the data}\label{sec:ci}

A \emph{confidence interval} gives a range of effect sizes that are reasonably compatible with the data, given the model. Its guarantee concerns the method, not any single interval: if a study were repeated many times and a 95\% interval calculated each time, about 95\% of those intervals would contain the true value (this property is called \emph{coverage}). A particular computed interval either contains the true value or it does not. The common statement ``there is a 95\% probability the true value lies in this interval'' is not what a confidence interval means; that is a Bayesian statement and needs a prior \citep{Hoekstra2014,Morey2016}. Some authors therefore prefer the names \emph{compatibility interval} or \emph{uncertainty interval} \citep{Gelman2019CI,Rafi2020}.

The guarantee breaks if intervals are reported selectively. In the Cochrane analysis of 23{,}551 trials, nominal 95\% intervals from trials reporting $p$ between 0.001 and 0.05 contained the true effect only about 90\% of the time, because selecting ``significant'' results favours overestimates \citep{vanZwet2024}.

\noindent\textit{In the reading-app study.} The 95\% interval from the 100-pupil trial runs from 0.02 to 0.58 standard deviations. In plain words, the data are compatible with anything from a trivial gain to a large one. Reporting only ``$p=0.034$, significant'' would hide that uncertainty.

\begin{mathbox}{Box 2. Confidence interval and coverage}
A $100(1-\alpha)\%$ confidence interval $C(X)$ satisfies $\Pr_\theta\{\theta\in C(X)\}\ge1-\alpha$ for every value of the parameter $\theta$, at a fixed design. For confidence sets constructed by inversion of a corresponding family of level-$\alpha$ tests, the confidence set consists exactly of the parameter values not rejected by those tests.
\end{mathbox}

\subsection{Equivalence tests: is the effect too small to matter?}\label{sec:tost}

A non-significant result (say $p=0.30$) does not show that there is no effect; the study may simply have been too small to tell. To claim that an effect is negligible, researchers first need to say how small an effect would be too small to care about. This threshold is the \emph{smallest effect size of interest} (SESOI), and it should be chosen and justified before the data are seen, on practical, clinical, or economic grounds \citep{Lakens2018Tutorial,Lakens2022}.

\begin{figure}[t]
\centering
\includegraphics[width=\textwidth]{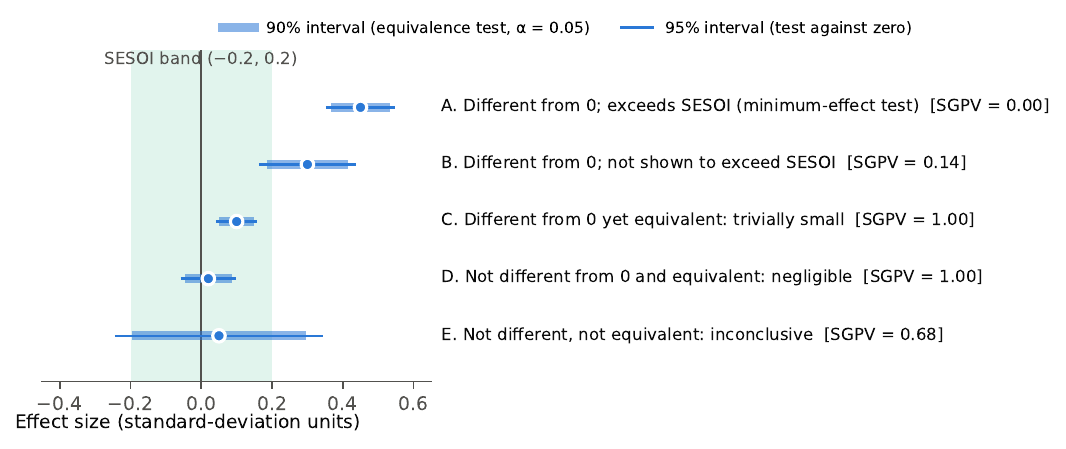}
\caption{Reading intervals against a smallest effect size of interest of $\pm0.2$ (shaded band). Thin lines are 95\% intervals, displayed for comparison with ordinary two-sided testing against zero; thick bars are 90\% intervals, which determine equivalence under TOST at $\alpha=0.05$. Patterns C and D are classified as equivalent because their 90\% intervals lie inside the band. The bracketed numbers are second-generation $p$-values (Section~\ref{sec:sgpv}) computed from the 95\% intervals.}
\label{fig:equiv}
\end{figure}

An \emph{equivalence test} then asks whether the data rule out effects as large as the SESOI in either direction. The standard version, the \emph{two one-sided tests} (TOST) procedure, amounts to checking whether a confidence interval lies entirely inside the band from $-$SESOI to $+$SESOI \citep{Schuirmann1987,Lakens2017}. Note which interval: an ordinary two-sided test against zero at $\alpha=0.05$ corresponds to the 95\% interval, whereas TOST at $\alpha=0.05$ corresponds to the $100(1-2\alpha)\%=90\%$ interval, because each of its two one-sided tests is performed at level $\alpha$. A related \emph{minimum-effect test} asks the opposite question: does the effect exceed the SESOI? Figure~\ref{fig:equiv} shows the five patterns that arise when these tests are combined with an ordinary test against zero. Only an equivalence test, or an equivalent interval-based method, can support a claim that an effect is negligible.

\noindent\textit{In the reading-app study.} The district set its SESOI at 0.2 before the study. Had the trial ended with $p=0.30$, the district could not conclude that ``the app does nothing''; to claim that any gain is too small to matter, it would need the 90\% interval to lie entirely between $-0.2$ and $+0.2$. In the 100-pupil trial the interval stretches well beyond 0.2, so the app is neither shown to be negligible nor shown to exceed the SESOI.

\begin{mathbox}{Box 3. Equivalence (TOST) and minimum-effect tests}
Let $\Delta>0$ be the SESOI and $\delta$ the true effect. Equivalence testing tests $H_0:|\delta|\ge\Delta$ against $H_1:|\delta|<\Delta$. TOST rejects $H_0$ at level $\alpha$ when both one-sided tests of $\delta\le-\Delta$ and $\delta\ge\Delta$ reject at level $\alpha$; equivalently, when the $100(1-2\alpha)\%$ confidence interval lies inside $(-\Delta,\Delta)$ \citep{Schuirmann1987}. A minimum-effect test tests $H_0:\delta\le\Delta$ against $H_1:\delta>\Delta$.
\end{mathbox}

\subsection{Second-generation $p$-values: how much of the interval is negligible?}\label{sec:sgpv}

The \emph{second-generation $p$-value} (SGPV) builds the SESOI directly into a single number \citep{Blume2018,Blume2019}. It is an overlap-based measure of compatibility between the interval estimate and the interval null of negligible effects. When the interval is sufficiently precise it reduces to the fraction of the interval lying inside the ``negligible'' band; when the interval is very wide, an additional correction prevents uninformative data from appearing strongly conclusive (Box~4). An SGPV of 0 means the interval lies entirely outside the band (the data are compatible only with effects that matter); 1 means it lies entirely inside (compatible only with negligible effects); values in between mean the data are inconclusive. Despite the name, the SGPV is not a $p$-value in the classical sense: it is not a probability, and its error properties come from the interval used to compute it. It has also been extended to choosing variables in models with many candidate predictors \citep{Zuo2022}.

\noindent\textit{In the reading-app study.} The 100-pupil trial gives an SGPV of 0.32: about a third of the plausible range of effects lies in the negligible band, so the data are inconclusive about whether the app matters.

\begin{mathbox}{Box 4. Second-generation $p$-value}
For an interval estimate $I$ and an interval null $H_0$ of negligible effects, with $|\cdot|$ denoting length,
\[ p_\delta=\frac{|I\cap H_0|}{|I|}\times\max\left\{\frac{|I|}{2|H_0|},\,1\right\}. \]
The second factor pulls $p_\delta$ towards $1/2$ when the interval is more than twice as wide as the null band, flagging uninformative data \citep{Blume2018}.
\end{mathbox}

\subsection{Bayes factors and Bayesian analysis}\label{sec:bayes}

Bayesian methods describe uncertainty with probabilities and update them as data arrive. Researchers begin with a \emph{prior}, a probability distribution expressing what was plausible before the study, and combine it with the data to obtain a \emph{posterior}, an updated distribution \citep{Gelman2013BDA}. A \emph{Bayes factor} compares two hypotheses by asking how much better one predicted the observed data than the other. A Bayes factor of 5 means the data were five times more likely under the first hypothesis than under the second \citep{KassRaftery1995}. Unlike a $p$-value, a Bayes factor can express evidence \emph{for} the null model \citep{Rouder2009,Keysers2020}. It is important, however, to keep apart several different claims about ``no effect'':
\begin{itemize}
\item \emph{evidence for an exact null}: a point-null Bayes factor compares $H_0:\delta=0$ with a nonzero-effect model;
\item \emph{practical equivalence}: the claim that $|\delta|<\Delta$, that is, any effect is smaller than the SESOI;
\item \emph{interval-null Bayes factors}, which compare $|\delta|<\Delta$ with $|\delta|\ge\Delta$ and so address practical equivalence in Bayesian terms;
\item \emph{TOST} (Section~\ref{sec:tost}), the frequentist test of practical equivalence;
\item the \emph{SGPV} (Section~\ref{sec:sgpv}), a descriptive overlap measure; and
\item \emph{ROPE-based posterior summaries}, which report how much posterior probability lies inside a \emph{region of practical equivalence} \citep{Kruschke2018ROPE}.
\end{itemize}
A point-null Bayes factor favouring $\delta=0$ does not by itself show practical equivalence, and an equivalence test does not show that the effect is exactly zero.

Bayes factors have two well-known sensitivities. First, they depend on the prior chosen for the effect under the alternative hypothesis, so priors must be justified and their influence checked. Second, for the same $p$-value, the Bayes factor drifts towards the null as the sample grows. This is the \emph{Jeffreys--Lindley paradox} \citep{Lindley1957}; Section~\ref{sec:worked} explains its mechanism and demonstrates both sensitivities.

\noindent\textit{In the reading-app study.} For illustration, let $H_0:\delta=0$ and let the alternative model assign $\delta\sim N(0,0.5^2)$. Because this prior is symmetric around zero, $H_1$ allows both positive and negative effects rather than representing the directional claim that the app helps; it is a \emph{nonzero-effect model}. The 100-pupil trial gives $\mathrm{BF}_{10}\approx2.2$: the observed data are about twice as likely under the nonzero-effect model as under the exact null. That is weak evidence, much weaker than ``$p=0.034$, significant'' suggests to many readers. With prior spreads of 0.2 and 1.0 instead of 0.5, the Bayes factor would be 2.6 and 1.3. A genuinely directional question---does the app \emph{help}?---would require a one-sided or truncated prior and different Bayes factors.

A common belief is that Bayesian analyses are unaffected by when a researcher decides to stop collecting data. The posterior itself does not depend on the stopping rule, but whether decisions based on Bayes factors keep a low false-alarm rate when researchers stop as soon as the evidence looks good depends on the prior and the hypotheses; it can fail for some widely used default priors \citep{deHeide2021,Hendriksen2021}. Sequential Bayesian designs are therefore best accompanied by simulations of their error rates \citep{Schonbrodt2017}. Modern \emph{Bayesian workflow} adds routine checks of the prior and model, computational diagnostics, and sensitivity analyses \citep{Gelman2020Workflow,Vehtari2017}, and the Bayesian Analysis Reporting Guidelines (BARG) describe how to report them \citep{Kruschke2021BARG}.

\begin{mathbox}{Box 5. Bayes factor}
For hypotheses $H_0$ and $H_1$ with prior densities $\pi_0$ and $\pi_1$ on the parameter $\theta$,
\[ \mathrm{BF}_{10}=\frac{p(x\mid H_1)}{p(x\mid H_0)}=\frac{\int p(x\mid\theta)\,\pi_1(\theta)\,d\theta}{\int p(x\mid\theta)\,\pi_0(\theta)\,d\theta},\qquad \frac{\Pr(H_1\mid x)}{\Pr(H_0\mid x)}=\mathrm{BF}_{10}\times\frac{\Pr(H_1)}{\Pr(H_0)}. \]
Frequentist error control under optional stopping holds for simple nulls with priors that do not depend on the sample size, and for certain composite nulls with special (for example, right-Haar) priors, but can fail otherwise \citep{Hendriksen2021}.
\end{mathbox}

\subsection{E-values: evidence you can monitor as data arrive}\label{sec:evalues}

Many studies are not analysed once at a fixed sample size. Online experiments are watched continuously, trials have interim looks, and researchers sometimes add participants when results are ``almost significant.'' Checking an ordinary $p$-value repeatedly and stopping when it dips below 0.05 (\emph{optional stopping}) raises the false-alarm rate well above 5\% (Section~\ref{sec:sim}).

\emph{E-values} were developed for exactly this situation \citep{Vovk2021,Shafer2021,Grunwald2024}. An e-value can be read as the winnings of a bettor who starts with one unit of money and bets against the null hypothesis in a game that is fair if the null is true. If the null is true, the bettor is unlikely to get rich; turning one unit into 20 happens with probability at most 1 in 20. A large e-value is therefore evidence against the null, and rejecting when the e-value reaches $1/\alpha$ (20 for $\alpha=0.05$) controls the false-alarm rate. When evidence is updated observation by observation (an \emph{e-process}), this control holds no matter when or why the researcher stops or continues, a property called \emph{anytime validity}. The interval version, a \emph{confidence sequence}, is a range that is valid at every sample size at once \citep{Howard2021,Ramdas2023SAVI}.

E-values connect to familiar ideas. For simple hypotheses, likelihood ratios and Bayes factors are e-values; e-values from independent studies can be multiplied, which suits cumulative or ``living'' meta-analyses \citep{terSchure2022}; and there is an e-value version of the false discovery rate procedure that works under any dependence between tests \citep{WangRamdas2022}. E-values also support certain forms of post hoc threshold selection while retaining explicit risk guarantees, provided decisions are made within the corresponding e-value decision framework \citep{Grunwald2024PNAS}.

\noindent\textit{In the reading-app study.} Suppose the district, keen for a quick answer, looks at the results after every 10 pupils and would stop as soon as $p<0.05$. In the simulation of Section~\ref{sec:sim}, with 100 interim looks, this habit produced false positives in 37.6\% of studies. If instead the district tracks an e-process and declares success only when it reaches 20, it can look as often as it likes and the false-positive probability is guaranteed not to exceed 5\%. The threshold 20 is $1/\alpha$ for $\alpha=0.05$; it is a rejection threshold with that guarantee, not an evidential equivalent of $p=0.05$.

Two cautions apply. First, anytime validity has a cost in power. A procedure that controls false alarms at every possible stopping time cannot, at a sample size fixed in advance, be more powerful than the best test designed for that sample size alone, and in practice it is less powerful. The \emph{size} of the cost is not fixed: it depends on the betting or mixture strategy, on how well it matches the true effect, and on how long monitoring continues (Section~\ref{sec:sim}). And the name clashes with the VanderWeele--Ding ``E-value,'' an unrelated measure of how strong unmeasured confounding would need to be to explain away an observed association \citep{VanderWeeleDing2017}. Papers should say which is meant.

\begin{mathbox}{Box 6. E-values, e-processes, and confidence sequences}
A non-negative statistic $E$ is an \emph{e-variable} for a null hypothesis $\mathcal{H}_0$ if $\mathbb{E}_P[E]\le1$ for every $P\in\mathcal{H}_0$. By Markov's inequality, $\Pr_P(E\ge1/\alpha)\le\alpha$. A non-negative process $(E_t)$ is an \emph{e-process} if $\mathbb{E}_P[E_\tau]\le1$ for every $P\in\mathcal{H}_0$ and every stopping time $\tau$; for a test supermartingale $(M_t)$, Ville's inequality gives $\Pr_P(\sup_t M_t\ge1/\alpha)\le\alpha$ \citep{Ramdas2023SAVI}. This is an upper bound, not an equality: a procedure may attain a false-positive probability well below $\alpha$. A confidence sequence satisfies $\Pr(\forall t:\theta\in C_t)\ge1-\alpha$ \citep{Howard2021}.

Conversions: $e=\kappa p^{\kappa-1}$ ($0<\kappa<1$) turns a $p$-value into an e-value, and $p=\min(1,1/e)$ turns an e-value into a valid $p$-value \citep{Vovk2021}. For a simple null, every likelihood ratio and every Bayes factor with a fixed prior is an e-variable; for composite nulls special constructions are needed \citep{Wasserman2020,Grunwald2024}. The e-BH procedure rejects the $k^\ast$ hypotheses with the largest e-values, $k^\ast=\max\{k:e_{(k)}\ge m/(\alpha k)\}$, and controls the false discovery rate at $\alpha$ under arbitrary dependence \citep{WangRamdas2022}.
\end{mathbox}

\subsection{Testing many hypotheses at once}\label{sec:mult}

When many hypotheses are tested---thousands of genes, dozens of outcomes, many subgroups---some will look ``significant'' by chance alone. Two kinds of correction are common. Controlling the \emph{family-wise error rate} (FWER), for example with the Bonferroni or Holm methods, limits the chance of even one false positive. Controlling the \emph{false discovery rate} (FDR) limits the expected \emph{proportion} of false positives among the reported discoveries, and is less strict \citep{BenjaminiHochberg1995}. The standard Benjamini--Hochberg (BH) procedure works when tests are independent or positively related; the Benjamini--Yekutieli correction or the e-value version (e-BH) works under any dependence \citep{BenjaminiYekutieli2001,WangRamdas2022}. Even after correction, the effects selected as discoveries tend to be overestimated (the ``winner's curse''), so their size should be confirmed in new data or adjusted by shrinkage \citep{GelmanCarlin2014,vanZwet2024}.

\noindent\textit{In the reading-app study.} If the district measures 20 outcomes---reading, spelling, vocabulary, each grade separately---and the app truly does nothing, two different quantities describe the risk of a spurious finding. The \emph{expected number} of false positives at $\alpha=0.05$ is $20\times0.05=1$; the \emph{probability of at least one} false positive (with independent tests) is $1-0.95^{20}\approx0.64$. Reporting only the ``significant'' outcome would be misleading; Section~\ref{sec:multsim} shows how multiple-testing corrections change this picture.

\begin{mathbox}{Box 7. Benjamini--Hochberg procedure}
With $m$ tests and ordered $p$-values $p_{(1)}\le\dots\le p_{(m)}$, reject the hypotheses with the $k^\ast$ smallest $p$-values, where $k^\ast=\max\{k:p_{(k)}\le k\alpha/m\}$. This controls the FDR at level $\alpha$ under independence or positive regression dependence \citep{BenjaminiHochberg1995}.
\end{mathbox}

\subsection{Planned interim analyses in trials}\label{sec:gsd}

Clinical trials often look at accumulating data a few times before the planned end, for example to stop early if a treatment is clearly beneficial or harmful. \emph{Group-sequential designs} make this safe by setting stricter thresholds at each planned look, so that the overall false-alarm rate stays at the chosen level \citep{Wald1945,OBrienFleming1979,LanDeMets1983,JennisonTurnbull2000}. They work only if the looks follow the plan; e-processes extend the protection to unplanned looks and open-ended monitoring. Regulatory guidance on adaptive trials requires such planning and accepts Bayesian designs whose error rates are documented by simulation \citep{ICHE20}.

\noindent\textit{In the reading-app study.} A legitimate way for the district to look early is to plan, before the study, a single mid-term check with a stricter threshold (for example, about $p<0.005$ at mid-term and about $p<0.048$ at the end, as in an O'Brien--Fleming design), which keeps the overall false-alarm rate at 5\%.

\begin{landscape}
\small
\setlength{\tabcolsep}{4pt}
\begin{longtable}{L{2.6cm} L{2.3cm} L{3.6cm} L{4.3cm} L{4.1cm} L{3.6cm}}
\caption{Inferential summaries compared through the CDG framework.}\label{tab:tools}\\
\toprule
\textbf{Summary} & \textbf{Claims supported} & \textbf{Guarantee} & \textbf{Breaks down when} & \textbf{Common misreading} & \textbf{Correct reading}\\
\midrule
\endfirsthead
\toprule
\textbf{Summary} & \textbf{Claims supported} & \textbf{Guarantee} & \textbf{Breaks down when} & \textbf{Common misreading} & \textbf{Correct reading}\\
\midrule
\endhead
\midrule \multicolumn{6}{r}{\textit{Continued on next page}}\\
\endfoot
\bottomrule
\endlastfoot
$p$-value (point null) & Existence (as evidence); Action (with justified $\alpha$) & False-alarm control under a fixed plan: $\Pr(p\le\alpha)=\alpha$ if the full null model holds & Peeking and stopping; many tests or selective reporting; assumption violations; analysis choices made after seeing data & ``Probability $H_0$ is true''; ``no effect'' when $p>0.05$; small $p$ = large effect & Degree of incompatibility between data and the full model\\
S-value $-\log_2 p$ & Existence & Same as $p$ & Same as $p$ & Treated as a new test with different validity & Bits of information against the model; additive scale\\
Confidence (compatibility) interval & Size; Existence; Negligibility (with SESOI) & Coverage $\ge1-\alpha$ under a fixed plan & Selective reporting; peeking and stopping; model misspecification; no randomisation, for causal claims & ``95\% probability $\theta$ is in this interval'' & Values compatible with data under the model; procedure covers in $\ge95\%$ of repetitions\\
TOST / minimum-effect test & Negligibility / Existence with relevance & Fixed-plan false-alarm control for $H_0:|\delta|\ge\Delta$ (or $\delta\le\Delta$) & SESOI chosen after data; peeking; selective reporting & Non-significance read as equivalence; SESOI treated as arbitrary & Evidence that effect is inside (or beyond) prespecified bounds\\
SGPV $p_\delta$ & Size, existence, negligibility (descriptive); discovery (ProSGPV) & Inherits from interval; no tail-probability guarantee at intermediate values & As for the interval; interval null set post hoc & Read as a classical $p$-value & Fraction of compatible values that are negligible\\
Bayes factor & Comparison; Existence; Negligibility (interval null or ROPE) & Coherence given prior; frequentist calibration only under conditions & Prior sensitivity; Lindley divergence at large $n$; default priors under peeking & ``Posterior odds'' without prior odds; immune to all stopping rules & Relative predictive performance of $H_1$ vs $H_0$ given priors\\
Posterior / credible interval & Size; Action (with loss) & Coherence given prior & Prior or model misspecification; no randomisation & Credible interval read as having frequentist coverage & Probability statements conditional on model and prior\\
E-value / e-process & Existence; Discovery (e-BH); Action with post hoc $\alpha$ & False-alarm control (e-value); anytime validity (e-process) & Complex null without valid construction; no randomisation; post hoc analysis choices & Confused with VanderWeele--Ding E-value; read as posterior odds & Evidence as betting wealth against $H_0$; reject when $E\ge1/\alpha$\\
Confidence sequence & Size under monitoring & Anytime validity: $\Pr(\forall t:\theta\in C_t)\ge1-\alpha$ & Model misspecification; no randomisation & Expected to be as narrow as fixed-$n$ CI & Interval valid at every sample size simultaneously\\
BH / BY / e-BH & Discovery & Multiple-testing control: FDR $\le\alpha$ (BH under independence/PRDS; BY, e-BH under arbitrary dependence) & Unreported tests; dependence violating BH assumptions & FDR read as probability a given discovery is false & Expected false fraction among discoveries\\
Group-sequential test & Existence; Action & Fixed-plan false-alarm control for planned schedule & Unplanned looks; overrunning & Naive estimates after early stopping read as unbiased & Valid decision; estimates need bias adjustment\\
\end{longtable}
\end{landscape}
\normalsize

\section{When does method choice change conclusions?}\label{sec:matter}

The original motivation for this review was the observation that the choice among $p$-values, Bayes factors, and e-values sometimes matters a great deal and sometimes hardly at all. The CDG framework predicts where: method choice matters whenever the methods carry different guarantees and the design threatens one of them, or whenever they answer different claims. We illustrate with a worked example and two simulations; all numbers are reproducible from the provided supplementary code.

\subsection{A worked example: the same $p$-value, two sample sizes}\label{sec:worked}

We now put numbers on the reading-app study. Scores are in standard-deviation units with a known standard deviation of 1, so the standard error of the difference with $n$ pupils per group is $\sqrt{2/n}$; tests use the normal distribution; the SESOI is 0.2; and the Bayes factor compares the exact null $\delta=0$ with the nonzero-effect model $\delta\sim N(0,\tau^2)$, $\tau=0.5$. Consider two versions of the study. In the \emph{district trial}, 100 pupils use the app and 100 do not, and app users score 0.30 standard deviations higher. In a \emph{nationwide trial}, 10{,}000 pupils are in each group and app users score 0.03 higher. The two results were chosen to give exactly the same $p$-value. Table~\ref{tab:worked} reports each summary; Supplementary Appendix~\ref{app:calc} shows every calculation step by step so that the numbers can be checked by hand.

\begin{table}[t]
\centering
\small
\caption{The reading-app study: one $p$-value, two sample sizes. Both trials give $z=2.12$. Assumptions: known SD $=1$, normal tests, SESOI 0.2; $\mathrm{BF}_{10}$ compares the nonzero-effect model $\delta\sim N(0,\tau^2)$ with $\delta=0$. Step-by-step calculations are given in Supplementary Appendix~\ref{app:calc}.}
\label{tab:worked}
\begin{tabular}{L{5.2cm} L{4.2cm} L{4.2cm}}
\toprule
\textbf{Summary} & \textbf{District trial: 100 pupils/group} & \textbf{Nationwide trial: 10{,}000/group}\\
\midrule
Estimate $\hat\delta$ (SE) & 0.300 (0.141) & 0.030 (0.014)\\
Two-sided $p$ (point null) & 0.034 & 0.034\\
S-value & 4.9 bits & 4.9 bits\\
95\% compatibility interval & 0.023 to 0.577 & 0.002 to 0.058\\
90\% interval (for TOST) & 0.067 to 0.533 & 0.007 to 0.053\\
TOST $p$ for $|\delta|\ge0.2$ & 0.76 (not equivalent) & $<0.001$ (equivalent)\\
Minimum-effect test $p$ for $\delta\le0.2$ & 0.24 & $>0.99$\\
SGPV ($H_0=[-0.2,0.2]$) & 0.32 (inconclusive) & 1.00 (only negligible effects compatible)\\
$\mathrm{BF}_{10}$, $\tau=0.5$ (main analysis) & 2.2 & 0.27 ($\mathrm{BF}_{01}=3.7$)\\
$\mathrm{BF}_{10}$, $\tau=0.2$ / $\tau=1.0$ (sensitivity) & 2.6 / 1.3 & 0.66 / 0.13\\
Upper bound $1/(-e\,p\ln p)$ & 3.2 & 3.2\\
E-value ($=\mathrm{BF}_{10}$; simple null) & 2.2; e-based $p=0.46$ & 0.27; e-based $p=1$\\
\bottomrule
\end{tabular}
\end{table}

\begin{figure}[t]
\centering
\includegraphics[width=\textwidth]{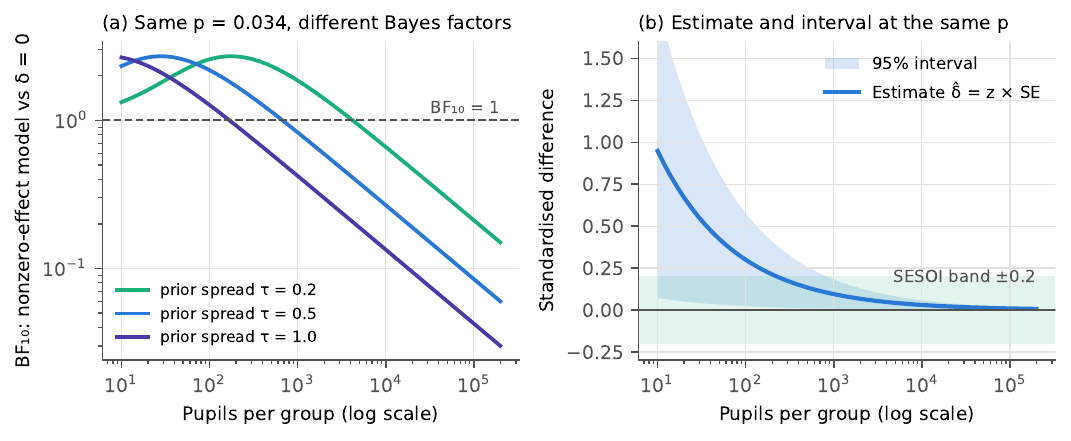}
\caption{The same $p$-value means different things at different sample sizes (the Jeffreys--Lindley paradox), illustrated with the reading-app study at $p=0.034$. (a) Bayes factor in favour of the specified nonzero-effect model $\delta\sim N(0,\tau^2)$ over $\delta=0$, for three prior spreads; it falls below 1 at about 173, 693, and 4{,}283 pupils per group for $\tau=1$, 0.5, and 0.2. (b) The estimate $\hat\delta=z\times\mathrm{SE}$ and its 95\% interval shrink into the band of negligible effects, so the same $p$-value accompanies an ever smaller gain.}
\label{fig:lindley}
\end{figure}

Three lessons follow. First, the $p$-value and S-value are identical across scenarios, but every summary that encodes magnitude or relevance changes. In the district trial, the data are compatible with gains from trivial to large: the app is statistically different from ``no effect'' but neither shown to be negligible nor shown to exceed the SESOI (pattern B in Figure~\ref{fig:equiv}). In the nationwide trial, the same $p$ accompanies a gain that is \emph{both} statistically different from zero \emph{and} clearly too small to be worth buying (pattern C). A headline reading ``reading app significantly improves scores ($p=0.034$)'' would be technically true in both cases and misleading in the second. Second, the Bayes factor moves from weak support for the nonzero-effect model to modest support for the exact null, a concrete instance of the Jeffreys--Lindley divergence. The mechanism is simple. At fixed $z$, the estimate is $\hat\delta=z\times\mathrm{SE}$, and because $\mathrm{SE}\propto n^{-1/2}$, holding $z$ fixed while $n$ grows forces $\hat\delta\to0$: the same $p$-value corresponds to an ever smaller observed effect. A diffuse alternative spreads its predictive probability over many effect sizes much larger than the one observed, while the point null concentrates its probability near zero, so the null eventually predicts the data better. Figure~\ref{fig:lindley}a shows that the crossover point depends strongly on the prior: at $p=0.034$, $\mathrm{BF}_{10}$ falls below 1 at about 173 pupils per group when $\tau=1$, 693 when $\tau=0.5$, and 4{,}283 when $\tau=0.2$. Third, the e-value derived from the same mixture likelihood ratio is far more conservative than $p$: an e-value of 2.2 does not reach the rejection threshold of 20 used for $\alpha=0.05$. Third, the e-value derived from the same mixture likelihood ratio is far more conservative than $p$: an e-value of 2.2 does not reach the threshold of 20 needed for rejection at $\alpha=0.05$.

The example corrects a claim sometimes made in reform discussions, including an earlier version of this manuscript: that in a large, preregistered fixed-sample trial all methods reach the same conclusion. They do not. A large sample removes stopping-rule and small-sample problems, but it sharpens the difference between asking whether there is any effect and asking whether the effect is large enough to matter, and it amplifies the divergence between $p$-values and Bayes factors. What the large preregistered trial does guarantee is that each summary is valid \emph{for its own claim}.

\subsection{Optional stopping: a simulation}\label{sec:sim}

To match the running example, the simulation reproduces a two-arm randomised trial. Pupils are enrolled in pairs, one to each group, and the district examines the accumulating data after every 10 pupils per group, up to 1000 per group (100 interim looks). Scores have standard deviation 1 in each group. We repeated this study $20{,}000$ times, first with no true effect ($\delta=0$) and then with $\delta=0.2$. We compared two monitoring rules: (i) stop and declare an effect at the first look at which an unadjusted two-sided $p<0.05$; and (ii) stop when an e-process reaches 20, which is the rejection threshold $1/\alpha$ that guarantees a false-positive probability of at most $\alpha=0.05$ (it is a threshold, not an evidential equivalent of $p=0.05$). The e-process is the normal-mixture likelihood ratio with prior spread $\tau$, that is, the Bayes factor of Section~\ref{sec:worked} recomputed at each look. To show how much the results depend on this choice, we report $\tau=0.2$, 0.5, and 1.0 (Box~8).

\begin{mathbox}{Box 8. Simulation details}
Pupils are enrolled in pairs; the within-pair difference $D_i=Y_{T,i}-Y_{C,i}\sim N(\delta,2)$. After $n$ pupils per group, $S_n=\sum_{i=1}^nD_i$ and $z_n=S_n/\sqrt{2n}$. Looks at $n=10,20,\dots,1000$; $R=20{,}000$ replications per scenario (seed 20261006). The naive rule rejects at the first look with $|z_n|>1.96$. The e-process is
\[ E_n=\sqrt{\frac{2}{2+n\tau^2}}\;\exp\!\left\{\frac{\tau^2S_n^2}{2\cdot2\,(2+n\tau^2)}\right\}, \]
a test martingale under $\delta=0$, so by Ville's inequality $\Pr_{H_0}(\sup_nE_n\ge20)\le0.05$. Monte Carlo standard errors are $\sqrt{\hat p(1-\hat p)/R}$.
\end{mathbox}

\begin{figure}[t]
\centering
\includegraphics[width=\textwidth]{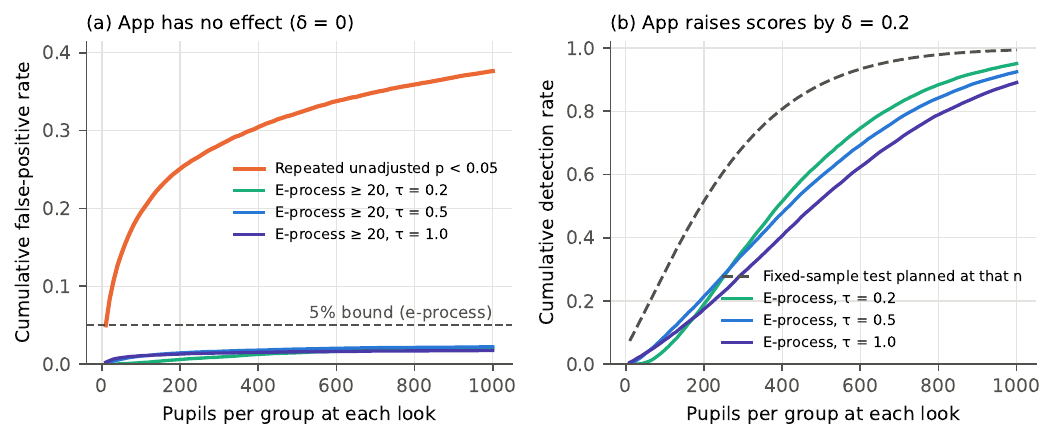}
\caption{Monitoring a two-arm reading-app trial after every 10 pupils per group (100 looks; $20{,}000$ simulated studies per scenario). (a) With no true effect, stopping at the first unadjusted $p<0.05$ produced false positives in 37.6\% of studies (Monte Carlo SE 0.34 points), a value specific to this monitoring scheme; the e-processes stayed near 2\%, below their theoretical 5\% upper bound. (b) With a true effect of 0.2, detection by the e-processes for three prior spreads $\tau$, compared with the power of a fixed-sample test planned at each sample size. Detection at a given sample size is lower than for a fixed-sample test planned at that size; some such loss is the general price of validity under unrestricted monitoring and stopping, and the prior spread changes its size but not its existence.}
\label{fig:sim}
\end{figure}

\paragraph{Results with no true effect.} With 100 interim looks between 10 and 1000 pupils per group, repeated application of an unadjusted $p<0.05$ rule produced a false-positive probability of 37.6\% (Monte Carlo standard error 0.34 percentage points; 95\% Monte Carlo interval 37.0\% to 38.3\%), and 19.5\% already by 100 pupils per group (Figure~\ref{fig:sim}a). These values are specific to this monitoring scheme: naive repeated testing has no fixed false-positive rate, and the inflation grows with the number and spacing of looks. A single fixed-sample test at 1000 per group rejected in 5.1\% (MC SE 0.16), as it should. The particular anytime-valid e-processes used here produced false-positive rates of 2.1\%, 2.2\%, and 1.8\% for $\tau=0.2$, 0.5, and 1.0 (MC SE about 0.1 percentage points each). The 5\% guarantee is an upper bound, not a requirement that the procedure attain exactly 5\%, and three features of this design make the procedure conservative. First, the bound from Ville's inequality covers monitoring that could continue forever, whereas this study stops at 1000 pupils per group. Second, the data are examined only every 10 pupils, so the e-process can rise above 20 and fall back between looks without being seen. Third, a mixture e-process spends part of its error budget on effect sizes far from those that occur, so under the null it rarely approaches the threshold. Together these explain why the observed rates are near 2\%.

\paragraph{Results with a true effect of 0.2.} With $\tau=0.5$, the e-process detected the effect in 21\% of studies by 200 pupils per group, 48\% by 400, and 92\% by 1000; among studies that stopped, the median stopping point was 390 pupils per group. For comparison, a fixed-sample test planned in advance at 200 pupils per group has power 0.52, and at 400 per group about 0.81 (Figure~\ref{fig:sim}b, dashed line). Detection by 1000 per group was 95\%, 92\%, and 89\% for $\tau=0.2$, 0.5, and 1.0, so the choice of mixture changes the size of the gap. It does not remove it: at 200 pupils per group, detection was between 17\% and 21\% for all three spreads, against 52\% for a fixed-sample test planned at that size. Some loss of power relative to a fixed-sample test is the general price of validity under unrestricted monitoring and stopping; how large it is depends on the construction. Group-sequential designs occupy a middle ground for a fixed, prespecified schedule of looks \citep{JennisonTurnbull2000}.

\subsection{Many outcomes: a multiple-testing simulation}\label{sec:multsim}

Suppose the district measures 20 outcomes, each compared between two groups of 100 pupils. We simulated $20{,}000$ such studies under two scenarios---the app has no effect on any outcome, and the app improves 3 of the 20 outcomes by 0.4 standard deviations (each test then has power 0.81 on its own)---and applied seven rules: unadjusted testing at $p<0.05$; Bonferroni and Holm (family-wise control); Benjamini--Hochberg (BH) at FDR 5\%, valid under independence or positive dependence; and three procedures valid under \emph{arbitrary} dependence between outcomes---Benjamini--Yekutieli (BY), and e-BH with two choices of e-value, the Bayes-factor e-values of Section~\ref{sec:worked} and e-values tuned in advance to the planning effect $\delta=\pm0.4$ (Table~\ref{tab:mult}).

\begin{table}[t]
\centering
\small
\caption{Twenty outcomes compared between two groups of 100 pupils ($20{,}000$ simulated studies per scenario; independent outcomes; $\alpha=0.05$ or FDR 5\%). ``P(at least one false positive)'' is the family-wise error rate; ``expected false positives'' is the average number. With 20 true nulls the theoretical values without correction are $1-0.95^{20}=0.642$ and $20\times0.05=1$.}
\label{tab:mult}
\begin{tabular}{L{3.5cm} L{1.9cm} L{1.9cm} L{1.9cm} L{1.3cm} L{2.2cm}}
\toprule
 & \multicolumn{2}{>{\centering\arraybackslash}p{4.2cm}}{\textbf{Scenario 1: no real effects}\newline (20 null outcomes)} & \multicolumn{3}{>{\centering\arraybackslash}p{6.0cm}}{\textbf{Scenario 2: three real effects}\newline (effect 0.4 on 3 outcomes; 17 nulls)}\\
\cmidrule(lr){2-3}\cmidrule(lr){4-6}
\textbf{Rule} & \textbf{P(at least one false positive)} & \textbf{Expected false positives} & \textbf{P(at least one false positive)} & \textbf{FDR} & \textbf{Real effects found (of 3)}\\
\midrule
Unadjusted & 0.644 & 1.00 & 0.583 & 0.218 & 2.43\\
Bonferroni & 0.047 & 0.05 & 0.042 & 0.022 & 1.27\\
Holm & 0.047 & 0.05 & 0.045 & 0.023 & 1.29\\
Benjamini--Hochberg (BH) & 0.048 & 0.05 & 0.109 & 0.042 & 1.54\\
Benjamini--Yekutieli (BY) & 0.012 & 0.01 & 0.026 & 0.012 & 1.02\\
e-BH, Bayes-factor e-values ($\tau=0.5$) & 0.002 & 0.00 & 0.002 & 0.001 & 0.43\\
e-BH, e-values tuned to $\delta=\pm0.4$ & 0.003 & 0.00 & 0.007 & 0.004 & 0.62\\
\bottomrule
\end{tabular}
\end{table}

Without correction, the probability of at least one false positive was 0.644 and the expected number was 1.00, matching the theoretical values; these answer different questions and both should be reported. Bonferroni and Holm held the probability of any false positive near 5\%, at the cost of finding only about 1.3 of the 3 real effects. BH kept the false discovery rate near 4\% while finding about 1.5. The three procedures that remain valid under any dependence between outcomes were all more conservative: BY found 1.0 real effects, and e-BH found 0.43 with the Bayes-factor e-values and 0.62 with e-values tuned to the planning effect. Tuning the e-values helps, but only modestly here. The main reason is structural rather than a poor choice of e-value: with 20 outcomes, e-BH needs an e-value of at least $20/0.05=400$ to make even one discovery, a far stricter requirement than BH's $p\le0.0025$, and e-BH buys robustness to arbitrary dependence that these independent outcomes do not need. When outcomes are independent or positively related, BH is the natural choice; BY or e-BH are worth their cost when dependence is unknown or adversarial, or when e-values arise naturally, for example from sequential monitoring. The choice of rule therefore changes which ``discoveries'' are reported, and the appropriate rule depends on whether the claim concerns any single false positive or the proportion of false discoveries.

\subsection{A typology of regimes}\label{sec:regimes}

Combining these demonstrations with the literature yields the following typology.

\paragraph{Method choice changes conclusions when}
\begin{enumerate}[label=(\roman*)]
\item \emph{researchers peek and decide when to stop} (the district checks weekly): ordinary fixed-sample $p$-values lose their calibration, while e-values, always-valid $p$-values, and planned interim analyses keep their guarantees, and Bayes factors keep it only under the conditions in Section~\ref{sec:bayes};
\item \emph{samples are large relative to the effect of interest} (a nationwide study): $p$-values and Bayes factors diverge, and ``is there any effect?'' and ``is the effect big enough to matter?'' get different answers (Table~\ref{tab:worked});
\item \emph{the claim is that an effect is negligible} (``the app makes no meaningful difference''): only methods that compare the result with the SESOI (equivalence tests, SGPV, interval-null Bayes factors, ROPE summaries) can support it; a point-null Bayes factor addresses a different claim, that the effect is exactly zero;
\item \emph{many outcomes are tested or results selected} (20 subjects and grades): unadjusted results lose their guarantees, and the choice of multiple-testing correction determines which discoveries are reported (Section~\ref{sec:multsim}, Table~\ref{tab:mult});
\item \emph{groups are not randomised and the claim is causal} (schools volunteered): the choice of which background factors to adjust for, and the sensitivity analysis, matter more than the choice of statistical summary \citep{HernanRobins2020,CinelliHazlett2020}.
\end{enumerate}

\paragraph{Method choice matters less when} the design is fixed and prespecified, the sample is moderate, assumptions are checked, and the signal is either clearly strong or clearly absent. In that regime, estimates with compatibility intervals, $p$-values, and Bayes factors with reasonable priors usually point in the same direction, and the choice can be guided by the audience and by the claim. Even here the summaries are not interchangeable: they answer different questions, and the interval or posterior should accompany any test.

\section{A method-selection matrix}\label{sec:matrix}

\begin{figure}[t]
\centering
\includegraphics[width=0.96\textwidth]{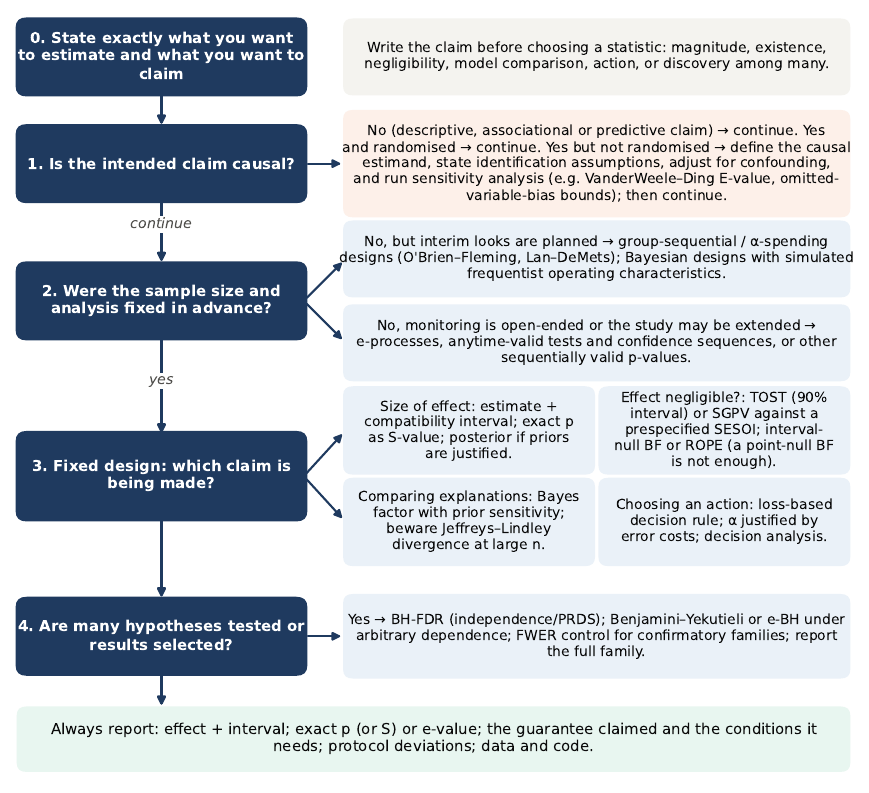}
\caption{Decision tree for choosing inferential summaries. Each branch corresponds to a feature of the study design or to the kind of claim being made.}
\label{fig:tree}
\end{figure}

Table~\ref{tab:matrix} maps common designs to the claims typically made, the summaries that license them, the guarantee required, the most frequent failure modes, and the reporting essentials. Figure~\ref{fig:tree} condenses the same logic into a decision tree. The matrix is a starting point, not a prescription: specialised settings (survival, cluster randomisation, diagnostic accuracy, measurement-error models) require domain-specific extensions.

\begin{landscape}
\small
\setlength{\tabcolsep}{3.5pt}
\begin{longtable}{L{2.7cm} L{2.0cm} L{3.4cm} L{3.0cm} L{2.4cm} L{3.4cm} L{3.6cm}}
\caption{Method-selection matrix: designs, claims, and the summaries that license them.}\label{tab:matrix}\\
\toprule
\textbf{Design / setting} & \textbf{Typical claim} & \textbf{Primary summary} & \textbf{Complements} & \textbf{Guarantee needed} & \textbf{Main failure modes} & \textbf{Reporting essentials}\\
\midrule
\endfirsthead
\toprule
\textbf{Design / setting} & \textbf{Typical claim} & \textbf{Primary summary} & \textbf{Complements} & \textbf{Guarantee needed} & \textbf{Main failure modes} & \textbf{Reporting essentials}\\
\midrule
\endhead
\midrule \multicolumn{7}{r}{\textit{Continued on next page}}\\
\endfoot
\bottomrule
\endlastfoot
Fixed-$n$ RCT, superiority & Size, Existence & Estimate + 95\% compatibility interval & Exact $p$ / S-value; minimum-effect test; posterior with justified prior & Coverage, Fixed-plan false-alarm control & Outcome switching; undisclosed subgroup emphasis & Estimand; prespecified analysis; all outcomes; CONSORT 2025\\
Fixed-$n$ RCT, decision (e.g.\ approval) & Action & Prespecified test at justified $\alpha$ & Estimate + interval; decision analysis & Fixed-plan false-alarm control & $\alpha$ not tied to error costs; post hoc thresholds & Justification of $\alpha$ and power; SPIRIT 2025 protocol\\
Equivalence / non-inferiority & Negligibility & TOST against prespecified margin; $(1-2\alpha)$ interval & SGPV; interval-null BF or ROPE & Fixed-plan false-alarm control for interval null & Margin chosen post hoc; non-significance read as equivalence & Margin justification; analysis population; both one-sided tests\\
Planned interim analyses & Existence, Action & Group-sequential / $\alpha$-spending test & Bias-adjusted estimates; Bayesian design with simulated operating characteristics & Fixed-plan false-alarm control (planned looks) & Unplanned looks; naive estimates after early stopping & Boundaries; information fractions; stopping reasons\\
Continuous or open-ended monitoring (A/B tests, living reviews) & Existence, Size & E-process; confidence sequence & Fixed-$n$ analysis at planned horizon; e-value meta-analysis & Anytime validity & Composite null without valid e-construction; reporting only the final look & E-process definition; threshold $1/\alpha$; full monitoring history\\
Replication study & Size, Negligibility & Estimate + interval vs original; equivalence to SESOI & Replication BF; small-telescopes or predictive checks & Coverage, Fixed-plan false-alarm control & Judging success by $p<0.05$ alone; ignoring original effect inflation & Original effect; SESOI; power for SESOI\\
Many hypotheses (omics, screening) & Discovery & BH-FDR (independence/PRDS); BY or e-BH (arbitrary dependence) & Effect-size thresholds; shrinkage; independent validation & Multiple-testing control & Unreported tests; dependence; winner's-curse estimates & Number of tests; procedure; validation results\\
High-dimensional regression ($p\gg n$) & Discovery, Calibrated prediction & Penalised regression with validation; ProSGPV & Selective-inference intervals; stability selection & Multiple-testing control or Calibrated prediction & Post-selection intervals treated as unadjusted & Tuning procedure; out-of-sample performance\\
Hierarchical / correlated data & Size & Multilevel model; posterior or mixed-model intervals & Prior sensitivity; posterior predictive checks & Coherence given prior or Coverage & Ignored clustering; unchecked priors & Model specification; priors; BARG items\\
Meta-analysis & Size & Random-effects estimate + interval and prediction interval & Bayesian hierarchical model; bias assessment & Coverage or Coherence given prior & Small-study effects; heterogeneity treated as noise & PRISMA 2020; heterogeneity; bias assessment\\
Observational causal study & Size (causal) & Adjusted estimate + interval for defined estimand & Sensitivity analysis (VanderWeele--Ding E-value; OVB bounds); negative controls & Coverage conditional on identification & Unmeasured confounding; selection bias & DAG; adjustment set; sensitivity results\\
Exploratory analysis & Hypothesis generation & Descriptive estimates; multiverse / specification curve & Holdout validation & None claimed & Exploratory results reported as confirmatory & Label as exploratory; report all analyses\\
\end{longtable}
\end{landscape}
\normalsize

\section{Integrating inference with the research workflow}\label{sec:workflow}

\begin{figure}[t]
\centering
\includegraphics[width=\textwidth]{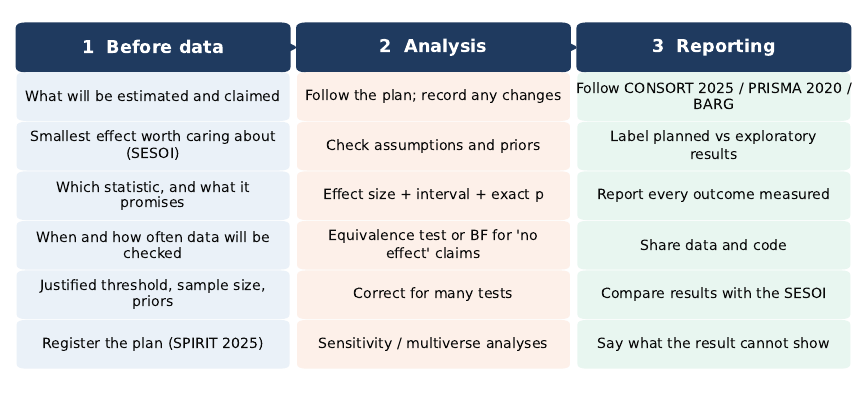}
\caption{A three-phase workflow that links the claim, design, and guarantee to reporting standards.}
\label{fig:workflow}
\end{figure}

The CDG framework implies that most of the work of valid inference happens before and after the analysis itself: before, by fixing the claim, the design, and the guarantee; after, by reporting enough for readers to check that the guarantee survived. Figure~\ref{fig:workflow} and Table~\ref{tab:workflow} organise this into three phases and map each element to current guidelines.

\begin{table}[t]
\centering
\small
\caption{Workflow elements, the CDG component they protect, and where current guidelines address them.}
\label{tab:workflow}
\begin{tabular}{L{4.4cm} L{2.0cm} L{3.0cm} L{4.6cm}}
\toprule
\textbf{Element} & \textbf{Protects} & \textbf{Phase} & \textbf{Guidance}\\
\midrule
Define estimand and claim type & C & Before data & ICH E9(R1); SPIRIT 2025\\
Justify SESOI, $\alpha$, sample size, priors & C, G & Before data & \citet{Lakens2018JustifyAlpha,Lakens2022}; BARG\\
Prespecify stopping rule or anytime-valid plan & Peeking and stopping; false-alarm control or anytime validity & Before data & SPIRIT 2025; ICH E20 (draft)\\
Register protocol / Registered Report & Many tests/selection, Post hoc analysis choices & Before data & SPIRIT 2025; \citet{Chambers2022}\\
Report deviations from plan & Post hoc analysis choices & Analysis/reporting & CONSORT 2025\\
Effect estimates with intervals; exact $p$ & Coverage, Fixed-plan false-alarm control & Analysis & SAMPL; CONSORT 2025\\
Interval-null procedure for absence claims & Negligibility & Analysis & \citet{Lakens2018Tutorial}; \citet{Keysers2020}\\
Multiplicity control matched to claim & Many tests/selection, Multiple-testing control & Analysis & SAMPL; \citet{BenjaminiHochberg1995}\\
Prior and model sensitivity analyses & Coherence given prior & Analysis & BARG; \citet{Gelman2020Workflow}\\
Multiverse / specification curve & Post hoc analysis choices & Analysis & \citet{Steegen2016,Simonsohn2020}\\
Label confirmatory vs exploratory & Post hoc analysis choices & Reporting & CONSORT 2025; PRISMA 2020\\
Share data and code & All & Reporting & CONSORT 2025 (open science section)\\
\bottomrule
\end{tabular}
\end{table}

\paragraph{Phase 1: before data.} Specify the \emph{estimand}---exactly what is to be estimated: in whom, for which outcome, how events such as dropping out of treatment are handled, and which summary (for example, a difference in means)---and the claim type. Define the SESOI and, if relevant, the equivalence margin. Choose the primary summary according to the claim and the anticipated design: a fixed-$n$ test and interval if the sample size is fixed; a group-sequential design for planned interim looks; an e-process or confidence sequence if monitoring will be continuous or the study may be extended. Justify $\alpha$, the sample size, and priors. Register the protocol or submit a Registered Report. \emph{Reading app:} the district writes down, before term starts, that the outcome is the end-of-term reading score, that a gain below 0.2 standard deviations is not worth the cost, how many pupils will take part, and whether and how it will look at results early; it then registers this plan.

\paragraph{Phase 2: analysis.} Follow the plan and log every deviation. Check assumptions and priors. Report the effect estimate, its compatibility interval, and the exact $p$-value (optionally as an S-value) or e-value. For claims of absence, use an interval-null procedure rather than a non-significant point-null test. Apply multiplicity control matched to the claim. Conduct sensitivity analyses and, where analytic choices are numerous, a multiverse or specification-curve analysis. \emph{Reading app:} the district reports the estimated gain with its interval, compares the interval with the 0.2 threshold, reports all outcomes it measured, and notes any departures from the plan, such as schools that dropped out.

\paragraph{Phase 3: reporting.} Follow the applicable guideline (CONSORT 2025, PRISMA 2020, BARG, SAMPL, or a discipline-specific equivalent). Distinguish confirmatory from exploratory findings, report all prespecified outcomes, share data and code where possible, interpret results against the SESOI and prior evidence, and state explicitly which guarantee is claimed and which design conditions it does not cover. \emph{Reading app:} the report says, for example, ``the app raised scores by 0.30 (95\% interval 0.02 to 0.58); we cannot yet tell whether the gain exceeds the 0.2 needed to justify the cost,'' rather than ``the app works ($p<0.05$).'' Common pitfalls at each phase, why they mislead, and how to avoid them are summarised in Supplementary Appendix~\ref{app:pitfalls} (Table~\ref{tab:pitfalls}).

\section{Discussion}\label{sec:discussion}

\subsection{Synthesis propositions}

We summarise the review in five propositions, each with its evidential basis (Table~\ref{tab:props}). These are claims about the literature, not new empirical findings, and they should be read with the limitations below.

\begin{proposition}[Design dominates]\label{prop:design}
Inferential validity is conditional on the design and analysis process. A method whose guarantee does not match the actual design may become invalid. Some design features, such as planned multiplicity or sequential monitoring, can be accommodated by methods constructed for those settings. Other problems, such as uncontrolled confounding, undisclosed selective reporting, or unrecorded analytic flexibility, cannot be repaired merely by substituting one statistical summary for another.
\end{proposition}

\begin{proposition}[Divergence is predictable]\label{prop:diverge}
Methods diverge systematically in identifiable regimes---optional stopping, large samples at fixed $p$, claims of negligibility, multiplicity, and non-randomised designs---so the statement that ``method choice does not matter'' is false in general, though approximately true in prespecified fixed designs with clear signals.
\end{proposition}

\begin{proposition}[Guidance has outpaced practice]\label{prop:practice}
Despite a decade of statements, the reporting of $p$-values has increased and the near-universal reporting of significant results has persisted.
\end{proposition}

\begin{proposition}[Anytime validity as common ground]\label{prop:savi}
E-values and e-processes provide a shared language in which likelihood ratios, certain Bayes factors, and frequentist error control coincide, making them a natural option where designs are sequential or open-ended.
\end{proposition}

\begin{proposition}[Standards are the lever]\label{prop:standards}
Reporting standards provide an implementation mechanism by making key design and analysis decisions visible, auditable, and increasingly expected by journals, reviewers, and regulators; they do not prescribe a single valid statistical method, but their 2025 revisions encode several of the decade's recommendations.
\end{proposition}

\begin{table}[t]
\centering
\small
\caption{Evidential basis of the synthesis propositions.}
\label{tab:props}
\begin{tabular}{L{1.6cm} L{6.4cm} L{2.1cm} L{4.2cm}}
\toprule
\textbf{Prop.} & \textbf{Principal evidence} & \textbf{Strength} & \textbf{Caveats}\\
\midrule
\ref{prop:design} & Many-analysts studies \citep{Silberzahn2018,BotvinikNezer2020,Breznau2022}; Registered Reports comparison \citep{Scheel2021}; optional-stopping and multiple-testing simulations (Sections~\ref{sec:sim}--\ref{sec:multsim}) & Strong & Field-specific samples; observational comparisons\\
\ref{prop:diverge} & Lindley divergence \citep{Lindley1957}; worked example (Table~\ref{tab:worked}); Bayesian optional stopping \citep{deHeide2021,Hendriksen2021} & Strong (mathematical) & Magnitudes depend on priors and effect sizes\\
\ref{prop:practice} & Text mining 1990--2025 \citep{Chavalarias2016,Choi2026}; Cochrane RCTs \citep{vanZwet2024} & Moderate & Biomedical focus; 2026 study is a preprint; text mining measures reporting, not reasoning\\
\ref{prop:savi} & \citet{Ramdas2023SAVI,Grunwald2024,RamdasWang2025,Chugg2026} & Moderate (theoretical) & Limited applied uptake; power cost\\
\ref{prop:standards} & \citet{Hopewell2025,Chan2025SPIRIT,ICHE20,Kruschke2021BARG} & Moderate & Adherence to guidelines is imperfect; effect on practice not yet measured\\
\bottomrule
\end{tabular}
\end{table}

\subsection{Barriers to adoption}

Several barriers explain the gap between guidance and practice. Incentives still reward positive, novel results \citep{Munafo2017,Scheel2021}. Many researchers learned NHST as a decision procedure and have little exposure to estimation, equivalence testing, or Bayesian methods; the 2016 guide to misinterpretations remains relevant precisely because the misreadings persist \citep{Greenland2016}. Newer tools have uneven software support: TOST, SGPV, and Bayes factors are implemented in widely used packages, whereas e-value methods are available for an expanding but still limited set of models (Supplementary Appendix~\ref{app:software}). Specifying a SESOI or a prior requires substantive judgement that many fields have not codified. Finally, journals and regulators vary in what they require; the 2025 revisions of SPIRIT and CONSORT and the ICH E20 draft may narrow this variation in clinical research, but comparable instruments are absent in many other fields.

\subsection{Limitations}

This is a narrative review. The search was targeted rather than exhaustive, and the selection of landmark contributions involves judgement; another author might weight Bayesian, likelihoodist, or severity-based perspectives \citep{Royall1997,Mayo2018} differently. The empirical evidence on practice is concentrated in biomedicine and psychology. The worked example uses a simple normal model with known variance and a particular prior; the qualitative conclusions are general, but the numbers are illustrative. The CDG framework is a conceptual tool; we have not evaluated whether it improves the quality of applied analyses, which would require a dedicated study.

\subsection{Future directions}

The decade settled many conceptual questions; the open questions now are largely empirical and practical. We see six priorities.

\paragraph{1. Evaluating the reforms themselves.} Most reforms of the decade have been adopted on the strength of argument rather than evidence of effect. We still know little about whether statements, journal policies, or the 2025 revisions of SPIRIT and CONSORT change what researchers actually report and conclude. Randomised or quasi-experimental evaluations of editorial and reporting interventions, and repeated text-mining studies of the kind reviewed here \citep{Chavalarias2016,Scheel2021}, are needed to tell which reforms work and which are symbolic.

\paragraph{2. Field-specific calibration of evidence.} The Cochrane analysis showed that the meaning of a $p$-value depends on the typical effect sizes and sample sizes of a field \citep{vanZwet2024}. Comparable empirical calibrations for education, psychology, economics, ecology, and other fields would allow researchers to report what a given result typically implies about exaggeration, sign errors, and replication, and would provide defensible default priors and shrinkage for individual studies \citep{GelmanCarlin2014}.

\paragraph{3. Making anytime-valid methods routine.} E-values and confidence sequences are mathematically mature for simple settings, but practitioners need methods for regression, clustered, longitudinal, and time-to-event data; tools for planning sample sizes; and guidance on reporting \citep{RamdasWang2025}. Their acceptance in confirmatory trials will depend on how the final ICH E20 guideline treats them \citep{ICHE20}, and their use in living systematic reviews needs empirical evaluation \citep{terSchure2022}.

\paragraph{4. Principled choice of thresholds and priors.} Equivalence tests and Bayesian analyses are only as good as the smallest effect size of interest and the priors behind them. Methods for eliciting these from stakeholders, field-level collections of justified SESOIs, and empirical checks of how sensitive published conclusions are to these choices would make relevance-based inference less subjective in practice \citep{Lakens2018Tutorial,Lakens2022}.

\paragraph{5. Measuring uncertainty from analysis choices.} Many-analysts studies show that analytic choices can change conclusions as much as sampling variation does \citep{Silberzahn2018,BotvinikNezer2020,Breznau2022}. Standards for when a multiverse or specification-curve analysis is expected, how it should be summarised, and how to combine it with ordinary uncertainty intervals remain to be developed \citep{Steegen2016,Simonsohn2020}.

\paragraph{6. Automation and AI-assisted analysis.} Automated tools can now check the internal consistency of reported statistics at scale \citep{Nuijten2016}, and AI systems increasingly assist with writing and running analyses. Such systems can make many analyses cheap to run and easy to discard, which magnifies the problems of peeking, multiplicity, and post hoc choices discussed in this review. Logging every analysis attempted, preregistering analysis code, and reporting the full set of analyses run will become more important, not less, and their effects deserve study.

Underlying all six is education. Statistics is still often taught as a menu of tests. Teaching each tool together with the question it answers and the conditions under which it can be trusted---and evaluating, through controlled studies, whether language such as compatibility intervals and S-values reduces misinterpretation \citep{Rafi2020}---may do more for research quality than any single new method.

\section{Conclusion}\label{sec:conclusion}

The decade from 2016 to 2026 is often described as a battle between $p$-values and their rivals. This review suggests a different reading. The statements, debates, and new methods of the period did not crown a winner. Their lasting contribution was to make explicit what had long been implicit: every inferential procedure carries an intended interpretation and, where applicable, a formal guarantee that is conditional on the design. An ordinary $p$-value guarantees rare false alarms if its sampling and analysis plan is respected; a confidence interval guarantees coverage if results were not selected; a Bayes factor is coherent if its prior is defensible; an e-process keeps its guarantee however often the data are examined. None of these guarantees extends to a confounded causal comparison or to an analysis chosen after the results were seen.

Seen this way, much of the controversy dissolves. Arguments over thresholds, over banning $p$-values, or over Bayesian versus frequentist methods were often arguments about different claims conducted in the vocabulary of a single statistic. The real dividing line is not between schools of statistics but between results whose conditions are stated and checked and results whose conditions are assumed and hidden. Methods matter, and we have shown where they lead to different answers, but they matter within a design, not instead of one.

This perspective also explains why the gap between guidance and practice has been so persistent. Better statistics cannot be adopted one number at a time; it requires changes in what researchers plan before collecting data, in what journals require authors to report, and in what students are taught to ask of a result. The infrastructure for this now exists in preregistration, Registered Reports, and the 2025 reporting standards. The remaining task is cultural: to expect, as a matter of routine, that every reported number arrives with the claim it supports and the conditions under which it can be trusted.

For the school district in our example, that would mean a report that reads not ``the app works ($p<0.05$)'' but ``the app raised reading scores by about 0.3 standard deviations, with plausible values from almost nothing to about 0.6; we cannot yet tell whether the gain exceeds what would justify the cost, and the study followed its registered plan.'' That sentence is longer, but it is honest about what is known, and it is the kind of sentence on which good decisions, in schools and in science, can be built.

\section*{Acknowledgements}
The authors thank [to be completed by the authors].

\section*{Funding}
This research did not receive any specific grant from funding agencies in the public, commercial, or not-for-profit sectors.

\section*{Disclosure statement}
The authors declare no conflict of interest.

\section*{Data and code availability}
No new empirical data were collected. The code (Python~3 with \texttt{numpy}, \texttt{scipy}, \texttt{matplotlib}) computes all entries in Table~3 (derived by hand in Appendix~\ref{app:calc}), the Bayes-factor prior-sensitivity results ($\tau=0.2$, 0.5, 1.0), the two-arm optional-stopping simulation of Section~6.2 (seed 20261006; $20{,}000$ replications per scenario; Monte Carlo standard errors reported), and the multiple-testing simulation of Section~6.3 (seed 20261007; $20{,}000$ replications per scenario), and draws Figures~3--6. It is publicly available on Kaggle at \url{https://www.kaggle.com/code/aswinipatra/statistics-simulation}.

\section*{Supplementary material}
The Supplementary Material contains: Appendix~A, software resources; Appendix~B, common pitfalls and remedies; Appendix~C, step-by-step calculations for Table~\ref{tab:worked}; Appendix~D, reproducible code; and Appendix~E, a glossary of terms.

\bibliography{references}

\clearpage
\input{supplementary}

\end{document}

%% file: supplementary.tex
\clearpage
\section*{Supplementary Material}

\appendix
\renewcommand{\thesection}{\Alph{section}}

\section{Software resources}\label{app:software}

Table~\ref{tab:software} lists freely available software for the tools described in Section~5 of the main article, grouped by method.

\begin{table}[H]
\centering
\small
\caption{Selected software for the methods discussed (versions change; consult current documentation).}
\label{tab:software}
\begin{tabular}{L{3.4cm} L{5.0cm} L{6.0cm}}
\toprule
\textbf{Method} & \textbf{Software} & \textbf{Notes}\\
\midrule
Estimation, intervals, tests & R (\texttt{stats}); Python (\texttt{scipy}, \texttt{statsmodels}) & Report exact $p$; S-value $=-\log_2 p$ computed directly\\
Equivalence / minimum-effect tests & R \texttt{TOSTER}; \texttt{statsmodels} (\texttt{ttost\_ind}); JASP & TOST, $(1-2\alpha)$ intervals, power for SESOI\\
Second-generation $p$-values & R \texttt{sgpv}; R \texttt{ProSGPV} & SGPV, delta-gaps, false discovery risk; variable selection\\
Bayes factors & R \texttt{BayesFactor}; \texttt{bayestestR}; JASP & Default and informed priors; ROPE summaries\\
Bayesian modelling and workflow & Stan (\texttt{cmdstanr}, \texttt{brms}); PyMC; \texttt{loo} & Prior/posterior predictive checks; LOO cross-validation\\
E-values, anytime-valid tests & R \texttt{safestats} & Safe $t$-tests, tests of two proportions, logrank tests\\
Confidence sequences & Python \texttt{confseq} & Time-uniform confidence sequences\\
Group-sequential designs & R \texttt{gsDesign}; R \texttt{rpact} & Boundaries, $\alpha$-spending, adaptive designs\\
Multiple testing & R \texttt{p.adjust}; \texttt{statsmodels} (\texttt{multipletests}) & BH, BY, Holm; e-BH is a short computation from sorted e-values\\
Sensitivity to confounding & R \texttt{EValue}; R \texttt{sensemakr} & VanderWeele--Ding E-value; omitted-variable-bias bounds\\
Multiverse / specification curve & R \texttt{multiverse}; R \texttt{specr} & Systematic enumeration of analytic choices\\
Preregistration & OSF Registries; ClinicalTrials.gov & Time-stamped protocols\\
\bottomrule
\end{tabular}
\end{table}

\section{Common pitfalls and remedies}\label{app:pitfalls}

Table~\ref{tab:pitfalls} summarises common errors in applying and interpreting the tools reviewed in the main article, why each misleads, and how to avoid it. It complements the workflow in Section~8 and the method-selection matrix in Table~5.

\begin{table}[H]
\centering
\small
\caption{Common pitfalls, why they mislead, and remedies, and the claim or design feature each one concerns.}
\label{tab:pitfalls}
\begin{tabular}{L{3.3cm} L{2.5cm} L{4.4cm} L{4.4cm}}
\toprule
\textbf{Pitfall} & \textbf{Concerns} & \textbf{Why it misleads} & \textbf{Remedy}\\
\midrule
Reading $p>0.05$ as ``no effect'' & Negligibility & A point-null test cannot support negligibility; wide intervals are uninformative & TOST or interval-null BF against a prespecified SESOI\\
Reading $p$ as $\Pr(H_0\mid\text{data})$ & Existence & $p$ is computed assuming $H_0$ and the full model & Report as compatibility or S-value; use BF/posterior if probability of hypotheses is wanted\\
Peeking until $p<0.05$ & Peeking/stopping & Type~I error rises with the number of looks (to 37.6\% with 100 looks in the simulation of Section~6.2; the value depends on the monitoring scheme) & Prespecify a group-sequential plan or use e-processes\\
Defining the SESOI after seeing data & Negligibility, Post hoc analysis choices & Margin can be tuned to the result & Prespecify and justify the SESOI in the protocol\\
Reporting only significant outcomes or subgroups & Many tests/selection & Selected estimates are inflated; intervals undercover & Report all outcomes; control FWER/FDR; label exploratory analyses\\
Treating a default prior as assumption-free & Coherence given prior & Bayes factors are sensitive to the prior under $H_1$ & Justify priors; report sensitivity analyses (BARG)\\
Claiming replication because both studies have $p<0.05$ & Size & Effects of very different size can both be significant & Compare estimates and intervals; test equivalence to the original or to the SESOI\\
Discovery without validation in high dimensions & Many tests/selection, Discovery & Winner's curse inflates selected effects & FDR control, effect-size thresholds, independent validation\\
Using a CI from an observational study as causal & No randomisation & Interval covers the associational parameter, not the causal one & Define causal estimand; adjust; sensitivity analysis for confounding\\
Confusing e-values with VanderWeele--Ding E-values & --- & Two unrelated quantities share a name & Define the quantity explicitly\\
\bottomrule
\end{tabular}
\end{table}

\section{Step-by-step calculations for Table~3}\label{app:calc}

This appendix shows how every entry in Table~3 is obtained, so that readers can check the numbers by hand or with a calculator. Values are rounded only at the final step; intermediate quantities are shown to four decimal places. Table~\ref{tab:calc} collects all intermediate quantities, and the numbered steps that follow explain how each is obtained.

\paragraph{Set-up and assumptions.} Reading scores are expressed in standard-deviation units, so the standard deviation in each group is taken as $\sigma=1$ and treated as known. With $n$ pupils per group, the difference in mean scores $\hat\delta$ has standard error
\[
\mathrm{SE}=\sqrt{\frac{\sigma^2}{n}+\frac{\sigma^2}{n}}=\sqrt{\frac{2}{n}}.
\]
All tests use the normal ($z$) distribution, with $\Phi$ denoting its cumulative distribution function. The smallest effect size of interest is $\Delta=0.2$, and the Bayes factor uses the prior $\delta\sim N(0,\tau^2)$ with $\tau=0.5$ under the alternative hypothesis.

\begin{table}[H]
\centering
\small
\caption{Intermediate quantities for the two scenarios in Table~3.}
\label{tab:calc}
\begin{tabular}{L{5.6cm} L{4.6cm} L{4.6cm}}
\toprule
\textbf{Quantity} & \textbf{District trial ($n=100$)} & \textbf{Nationwide trial ($n=10{,}000$)}\\
\midrule
Standard error $\sqrt{2/n}$ & 0.1414 & 0.0141\\
Estimate $\hat\delta$ & 0.3000 & 0.0300\\
$z=\hat\delta/\mathrm{SE}$ & 2.1213 & 2.1213\\
Two-sided $p=2\{1-\Phi(z)\}$ & 0.0339 & 0.0339\\
95\% interval $\hat\delta\pm1.96\,\mathrm{SE}$ & 0.0228 to 0.5772 & 0.0023 to 0.0577\\
TOST: $z_{\text{lower}}=(\hat\delta+\Delta)/\mathrm{SE}$ & 3.5355 ($p=0.0002$) & 16.26 ($p<10^{-50}$)\\
TOST: $z_{\text{upper}}=(\hat\delta-\Delta)/\mathrm{SE}$ & 0.7071 ($p=0.7602$) & $-12.02$ ($p<10^{-30}$)\\
TOST $p$ (larger of the two) & 0.7602 & $<0.001$\\
90\% interval $\hat\delta\pm1.645\,\mathrm{SE}$ (TOST) & 0.0674 to 0.5326 & 0.0067 to 0.0533\\
Minimum-effect $p=1-\Phi(z_{\text{upper}})$ & 0.2398 & $>0.99$\\
SGPV overlap $|I\cap H_0|$ & 0.1772 & 0.0554\\
SGPV interval width $|I|$ & 0.5544 & 0.0554\\
$|I|/(2|H_0|)$ (correction applies if $>1$) & 0.6930 (no correction) & 0.0693 (no correction)\\
SGPV $p_\delta$ & 0.3196 & 1.0000\\
$\tau^2/\mathrm{SE}^2$ & 12.5 & 1250\\
$\mathrm{BF}_{01}$ & 0.4575 & 3.7346\\
$\mathrm{BF}_{10}=1/\mathrm{BF}_{01}$ & 2.1858 & 0.2678\\
$\mathrm{BF}_{10}$ with $\tau=0.2$ (sensitivity) & 2.5875 & 0.6618\\
$\mathrm{BF}_{10}$ with $\tau=1.0$ (sensitivity) & 1.2712 & 0.1341\\
\bottomrule
\end{tabular}
\end{table}

\paragraph{Step 1: $z$ statistic and $p$-value.} For the district trial, $\mathrm{SE}=\sqrt{2/100}=0.1414$ and $z=0.30/0.1414=2.1213$. The two-sided $p$-value is $2\{1-\Phi(2.1213)\}=0.0339$. The nationwide trial was constructed so that $\hat\delta=z\times\mathrm{SE}=2.1213\times0.0141=0.0300$, giving the same $z$ and therefore the same $p$.

\paragraph{Step 2: S-value.} $s=-\log_2(0.0339)=4.88$ bits in both trials.

\paragraph{Step 3: 95\% compatibility interval.} District trial: $0.30\pm1.96\times0.1414=0.30\pm0.2772$, that is, 0.0228 to 0.5772. Nationwide trial: $0.03\pm1.96\times0.0141=0.03\pm0.0277$, that is, 0.0023 to 0.0577. (Using the rounded standard error 0.141 gives 0.024 to 0.576 for the district trial; the difference is rounding only.)

\paragraph{Step 4: equivalence test (TOST).} TOST performs two one-sided tests against the boundaries $\pm\Delta$ and reports the larger $p$-value. For the district trial, the test of $\delta\le-0.2$ gives $z=(0.30+0.20)/0.1414=3.5355$ and $p=0.0002$; the test of $\delta\ge0.2$ gives $z=(0.30-0.20)/0.1414=0.7071$ and $p=\Phi(0.7071)=0.7602$. The TOST $p$-value is therefore 0.76, and equivalence is not established. For the nationwide trial, both one-sided $p$-values are far below 0.001 ($z=16.26$ and $z=-12.02$), so the effect is shown to lie within $\pm0.2$. Equivalently, the 90\% interval ($\hat\delta\pm1.645\,\mathrm{SE}$)---not the 95\% interval, because each one-sided test is performed at $\alpha=0.05$---lies inside $(-0.2,0.2)$ for the nationwide trial (0.0067 to 0.0533) but not for the district trial (0.0674 to 0.5326).

\paragraph{Step 5: minimum-effect test.} This tests whether the effect exceeds the SESOI ($H_0:\delta\le0.2$). District trial: $z=(0.30-0.20)/0.1414=0.7071$, one-sided $p=1-\Phi(0.7071)=0.2398$, so the data do not show that the gain exceeds 0.2. Nationwide trial: $z=(0.03-0.20)/0.0141=-12.02$, $p>0.99$.

\paragraph{Step 6: second-generation $p$-value.} With $I$ the 95\% interval and $H_0=[-0.2,0.2]$ (width 0.4): for the district trial the overlap is $0.2-0.0228=0.1772$ and the interval width is $0.5772-0.0228=0.5544$. Because $|I|/(2|H_0|)=0.5544/0.8=0.693$ is below 1, no width correction applies, and $p_\delta=0.1772/0.5544=0.32$. For the nationwide trial the whole interval lies inside $H_0$, so the overlap equals the width and $p_\delta=1$.

\paragraph{Step 7: Bayes factor.} The Bayes factor compares the exact null $H_0:\delta=0$ with a \emph{nonzero-effect model} $H_1:\delta\sim N(0,\tau^2)$. Because this prior is symmetric around zero, $H_1$ allows both positive and negative effects; $\mathrm{BF}_{10}$ is therefore not evidence that the app \emph{helps}, which would require a one-sided or truncated prior. Under $H_0:\delta=0$, $\hat\delta\sim N(0,\mathrm{SE}^2)$; under $H_1$ with prior $\delta\sim N(0,\tau^2)$, $\hat\delta\sim N(0,\mathrm{SE}^2+\tau^2)$. The ratio of these two normal densities at the observed $\hat\delta$ is
\[
\mathrm{BF}_{01}=\sqrt{1+\frac{\tau^2}{\mathrm{SE}^2}}\;\exp\!\left\{-\frac{z^2}{2}\cdot\frac{\tau^2}{\mathrm{SE}^2+\tau^2}\right\}.
\]
District trial: $\tau^2/\mathrm{SE}^2=0.25/0.02=12.5$, so $\sqrt{13.5}=3.6742$; $\tau^2/(\mathrm{SE}^2+\tau^2)=0.25/0.27=0.9259$; $z^2/2=2.25$; hence $\mathrm{BF}_{01}=3.6742\times e^{-2.0833}=3.6742\times0.1245=0.4575$ and $\mathrm{BF}_{10}=2.19$. Nationwide trial: $\tau^2/\mathrm{SE}^2=1250$, $\sqrt{1251}=35.37$, $\tau^2/(\mathrm{SE}^2+\tau^2)=0.9992$; hence $\mathrm{BF}_{01}=35.37\times e^{-2.2482}=35.37\times0.1056=3.73$ and $\mathrm{BF}_{10}=0.27$. The large factor $\sqrt{1+\tau^2/\mathrm{SE}^2}$ in the nationwide trial is what drives the Jeffreys--Lindley divergence. Repeating the calculation with $\tau=0.2$ and $\tau=1.0$ gives $\mathrm{BF}_{10}=2.59$ and $1.27$ for the district trial and $0.66$ and $0.13$ for the nationwide trial, which shows how strongly the conclusion depends on the prior spread.

\paragraph{Step 8: upper bound on the Bayes factor.} With $p=0.0339$ and $\ln p=-3.3845$: $-e\,p\ln p=2.7183\times0.0339\times3.3845=0.3118$, so $\mathrm{BF}_{10}\le1/0.3118=3.21$ in both trials. The district Bayes factor (2.19) is below this bound, as it must be.

\paragraph{Step 9: e-value.} Because $H_0:\delta=0$ is a simple hypothesis (with $\sigma$ known) and the prior under $H_1$ is fixed in advance, $\mathrm{BF}_{10}$ is a valid e-value (Box~6). The e-values are therefore 2.19 and 0.27. The corresponding e-based $p$-values are $\min(1,1/e)$: $1/2.19=0.46$ for the district trial and 1 for the nationwide trial. Neither e-value approaches the rejection threshold $1/\alpha=20$, which guarantees a false-positive probability of at most 0.05 but is not an evidential equivalent of $p=0.05$.

\paragraph{Caveat.} These calculations assume normally distributed scores with known standard deviation. In a real analysis the standard deviation would be estimated and a $t$ distribution used; with 100 pupils per group the results would change only in the third decimal place.


\section{Glossary}\label{app:glossary}

\begin{description}[style=nextline,leftmargin=1.2cm,itemsep=3pt]
\item[Anytime-valid] A guarantee that holds however often the data are checked and whenever the study stops.
\item[Bayes factor] How many times more likely the observed data are under one hypothesis or model than under another; it can favour either. A point-null Bayes factor with a symmetric prior compares ``exactly zero'' with ``some nonzero effect,'' not ``helps'' with ``does nothing.''
\item[Claim types] The six kinds of claim a result can support: estimating a size, showing an effect exists, showing it is negligible, comparing explanations, choosing an action, and picking discoveries from many tests.
\item[Compatibility interval] Another name for a confidence interval, emphasising that it lists values compatible with the data under the model.
\item[Confidence interval] A range of effect sizes consistent with the data; the method captures the true value in a stated share (e.g.\ 95\%) of repeated studies.
\item[Confidence sequence] A sequence of intervals, one per sample size, that together contain the true value with the stated probability at all times.
\item[Confounding] Distortion of an apparent effect by a third factor that influences both the exposure and the outcome.
\item[Design features] Aspects of how a study is run that affect what results can be trusted: a fixed plan, peeking and stopping, many tests and selective reporting, no randomisation, and analysis choices made after seeing the data.
\item[E-value (e-variable)] A measure of evidence against the null that can be read as betting winnings; valid even under repeated checking when used as an e-process.
\item[E-value (VanderWeele--Ding)] An unrelated measure of how strong unmeasured confounding would need to be to explain away an association.
\item[E-process] A non-negative process $(E_t)$ that remains a valid e-value at any stopping time: $\mathbb{E}_P[E_\tau]\le1$ for every stopping time $\tau$ and every distribution $P$ in the null hypothesis. It can be updated as each new observation arrives and examined at any time.
\item[Effect size] The magnitude of a difference or association, such as a difference in means.
\item[Equivalence test (TOST)] A test of whether an effect is smaller than the SESOI in both directions.
\item[Estimand] The precise quantity a study aims to estimate (ICH E9(R1)).
\item[False discovery rate (FDR)] The expected share of false positives among all reported discoveries.
\item[Family-wise error rate (FWER)] The probability of at least one false positive among a set of tests.
\item[Guarantees] The promises a result carries: coverage, false-alarm control under a fixed plan, anytime validity, multiple-testing control, coherence given the prior, and calibrated prediction.
\item[Jeffreys--Lindley paradox] For the same $p$-value, the Bayes factor increasingly favours the null as the sample grows.
\item[Meta-analysis] Statistical combination of results from several studies.
\item[Meta-research] Research on how research is designed, conducted, and reported.
\item[Multiverse analysis] Reporting results under all reasonable combinations of analysis choices.
\item[Null hypothesis] The default assumption being tested, usually ``no effect.''
\item[Optional stopping] Deciding when to stop collecting data based on the results seen so far.
\item[$p$-value] The probability of data at least as extreme as those observed, assuming the null hypothesis and all other analysis assumptions are true.
\item[Posterior / prior] In Bayesian analysis, the prior expresses plausibility before the data; the posterior combines it with the data.
\item[Power] The probability that a study detects an effect of a given size if it exists.
\item[Preregistration] Publicly recording hypotheses and analysis plans before data collection.
\item[Randomised trial] A study in which chance decides who receives which treatment, so groups are comparable on average.
\item[Registered Report] A publication format in which a journal accepts a study based on its plan, before results are known.
\item[Replication] Repeating a study with new data to see whether its findings hold.
\item[ROPE] Region of practical equivalence; the Bayesian counterpart of the SESOI band.
\item[SESOI] Smallest effect size of interest; the threshold below which effects are considered too small to matter.
\item[SGPV] Second-generation $p$-value; the share of the confidence interval lying inside the negligible band, adjusted for interval width.
\item[Statistical significance] The conventional label for $p$ below a threshold (usually 0.05); it does not mean important or proven.
\item[S-value] $-\log_2 p$; the surprise in the data, in bits (coin tosses).
\item[Type I / Type II error] A false positive (declaring an effect that is not there) / a false negative (missing a real effect).
\end{description}